\documentclass[sn-mathphys-ay]{sn-jnl}

\usepackage{graphicx}
\usepackage{multirow}
\usepackage{amsmath,amssymb,amsfonts}
\usepackage{amsthm}
\usepackage{mathrsfs}
\usepackage[title]{appendix}
\usepackage{xcolor}
\usepackage{textcomp}
\usepackage{manyfoot}
\usepackage{booktabs}
\usepackage{algorithm}
\usepackage{algorithmicx}
\usepackage{algpseudocode}
\usepackage{listings}
\usepackage[normalem]{ulem}
\usepackage{xcolor}
\usepackage[figuresright]{rotating}
\theoremstyle{thmstyleone}

\theoremstyle{thmstyletwo}

\theoremstyle{thmstylethree}

\usepackage{comment}
\allowdisplaybreaks

\newcommand{\SP}[1]{{#1}}

\newcommand{\holdingCosts}{2}
\newcommand{\holdingCostsWIP}{1}

\newcommand{\holdingCostsComponents}{1}
\newcommand{\holdingCostsWIPComponents}{0.5}
\newcommand{\demandInformationHorizon}{12}

\newcommand{\Rev}[1]{{#1}}

\begin{document}

\title{Enhancing Rolling Horizon Production Planning Through Stochastic Optimization Evaluated by Means of Simulation}

\author[1]{\fnm{Manuel} \sur{Schlenkrich}}\email{manuel.schlenkrich@jku.at}

\author*[2]{\fnm{Wolfgang} \sur{Seiringer}}\email{wolfgang.seiringer@fh-steyr.at}

\author[2]{\fnm{Klaus} \sur{Altendorfer}}\email{klaus.altendorfer@fh-steyr.at}

\author[1]{\fnm{Sophie N.} \sur{Parragh}}\email{sophie.parragh@jku.at}

\affil[1]{Institute of Production and Logistics Management, Johannes Kepler University Linz, Altenbergerstra\ss e 69, 4040 Linz, Austria}

\affil[2]{Department for Production and Operations Management, University of Applied Sciences Upper Austria, Wehrgrabengasse 1, 4400 Steyr, Austria}

\abstract{
Production planning must account for uncertainty in a production system, arising 
from fluctuating demand forecasts \textcolor{black}{and execution-level friction}. 
\textcolor{black}{This article integrates} scenario-based stochastic programming 
\textcolor{black}{into} a rolling horizon \textcolor{black}{framework for} capacitated 
lot sizing, \textcolor{black}{evaluated via} discrete-event simulation. We 
\textcolor{black}{compare this stochastic approach against} deterministic optimization 
and standard Material Requirements Planning (MRP) \textcolor{black}{across varying} 
customer \textcolor{black}{update} behaviors, \textcolor{black}{shop loads, and diverse 
multi-stage topologies (divergent, convergent, and mixed). To accurately capture 
shop-floor dynamics, the framework introduces a non-anticipativity parameter 
controlling schedule flexibility, alongside probabilistic setup-time feedback 
and soft overtime constraints.}

\textcolor{black}{Results indicate that optimization consistently outperforms MRP. 
In unbuffered, highly congested settings, stochastic optimization natively smooths 
workloads and reduces costs by up to 68\%. However, introducing explicit safety 
stocks fundamentally shifts system dynamics: physical buffers effectively absorb 
shop-floor noise, diminishing the stochastic model's anticipative advantage and 
enabling deterministic optimization to dominate. Ultimately, this study offers 
critical managerial insights for aligning planning algorithms, inventory buffering, 
and schedule flexibility.}
}


\maketitle

\section{Introduction}
Fluctuating demand forecasts complicate supply chain management in various industrial sectors, necessitating medium-term production planning to mitigate customer driven uncertainty. In these dynamic environments, effective medium-term production planning requires rolling horizon planning and regular updates of customer demand, inventory levels, and available capacity  \citep{Stefansson2009}. Business customers typically provide \textcolor{black}{medium-term }demand forecasts \textcolor{black}{several weeks to a few months before the due date. These forecasts} are regularly adjusted to ensure \textcolor{black}{demand anticipation}, a practice known as  forecast evolution and necessitates dynamic adjustments of production plans \citep{Heath.1994, Norouzi.2014}.

\textcolor{black}{This behavior occurs across B2B supply chains, including 
automotive, industrial equipment, semiconductor, and contract manufacturing. 
Customers exhibit varying forecast update patterns: some make marginal 
adjustments, others revise regularly, and some delay substantial changes 
until just before the due date. Since each pattern distinctly impacts 
planning stability, they require tailored parameterizations of the 
medium-term planning method.} 

While medium-term planning can use deterministic or stochastic mathematical optimization models, industry often employs simpler approaches like material requirements planning (MRP) \citep{Orlicky.1975, Hopp.2011}.  
\textcolor{black}{Despite the pervasiveness of MRP, it often struggles to account for the inherent volatility of production environments. Consequently,} stochastic optimization is well-suited for dynamic settings as it incorporates uncertainties into the decision-making process, enhancing the robustness and adaptability of production plans.

A suitable concept to implement the described problem domain in one application framework is the combination of discrete-event simulation with an optimization solver. Simulation provides a controlled environment for assessing various production strategies and their potential outcomes before operational implementation, enhancing decision-making and resource utilization \citep{Negahban.2014}. \textcolor{black}{Simulation accommodates various operational contingencies by incorporating stochastic demand alongside shop-floor disruptions, such as volatile setup times, machine failures, or yield fluctuations}  \citep{Law.2014}, thereby helping to increase the overall efficiency and performance of manufacturing systems \citep{Sahin.2013}. The optimization component solves an optimization problem based on the current system status, incorporating stochastic parameters, like customer demands \textcolor{black}{or setup time variation}.

Other researchers have demonstrated the potential of simulation-optimization for medium-term planning. \citet{gansterer_simulation-based_2014} utilized simulation-optimization to determine good planning parameters for MRP under four distinct demand scenarios. \citet{almeder2009simulation} combined discrete-event simulation and optimization to identify an operation plan for a supply chain network based on a case study. In \cite{grunow_dynamic_stochastic_2023}, rolling-horizon planning and forecasting are integrated into a capacitated lot-sizing formulation, and numerical results are computed using simulation.

\textcolor{black}{While these foundational studies demonstrate the value of simulation-optimization, they primarily focus on parameterizing existing MRP frameworks or addressing isolated case studies. In contrast, to overcome the structural limitations of traditional MRP --- namely its reliance on static parameters and infinite-capacity assumptions under volatility --- this work investigates whether a unified, simulation-coupled stochastic optimization framework can entirely replace the MRP paradigm across diverse production topologies.} 

Our objective is to minimize overall costs in multi-item, multi-stage production systems, encompassing inventory and backlog costs. Through simulation, we benchmark deterministic and stochastic optimization against standard MRP, highlighting the potential effectiveness of the optimization results.

The theoretical contributions of this article are fourfold:

\begin{itemize}
    \item \textbf{Customer update behaviors and utilization:} Investigating distinct forecast evolution patterns alongside varying shop loads to derive managerial insights for planning under stochastic demand.
    \item \textbf{Evaluation of planning strategies:} Using discrete-event simulation to benchmark MRP against deterministic and stochastic optimization under shop-floor uncertainty, scaling from simple to complex multi-stage topologies.
    \item \textbf{Runtime behavior for scenario sampling:} Analyzing the impact of sample size in stochastic optimization to identify the optimal trade-off between computational efficiency and solution accuracy.
    \item \textcolor{black}{\textbf{Integrated capacity control:} Extending the optimization model with probabilistic setup-time feedback and endogenous overtime to dynamically balance planned safety buffers with execution-level recourse.}
\end{itemize}

The practical contributions are twofold: (1) \textcolor{black}{Enhancing} decision-making by integrating discrete-event simulation with stochastic optimization, providing a practical framework that helps managers make more informed decisions under uncertain demand. (2) \textcolor{black}{Improving} resource utilization by optimizing production schedules, thereby reducing overall production costs in dynamic manufacturing environments.

\textcolor{black}{This study addresses the following research questions:}  

\begin{itemize}
\item \textcolor{black}{RQ1:} How does the performance of stochastic optimization compare to deterministic optimization and standard MRP in rolling horizon production planning?
\item \textcolor{black}{RQ2:} What role does the number of scenarios in stochastic optimization play in achieving cost-effective production plans?
\item \textcolor{black}{RQ3:} How do different customer demand update behaviors influence the robustness and adaptability of production plans?
\item \textcolor{black}{RQ4:} How does varying the shop-floor load and production system complexity impact the performance of deterministic and stochastic optimization models?
\item  \textcolor{black}{RQ5: What is the effect of feeding execution-level setup time variability back into the optimization model alongside flexible overtime capacity?}
\end{itemize}

\section{Literature Review}
\label{sec:literature_review}
\subsection{Rolling Horizon Planning in Single- and Multi-Item Environments}
Several works consider single-item production environments, e.g. \cite{campuzano-bolarin_rolling_2020} investigate variable lead times for a single-item single-echelon system. They evaluate lot sizing techniques, namely Silver-Meal and Wagner-Within, in a system dynamics simulation and propose a rolling horizon planning approach. Also \cite{vargas_master_2011} study rolling horizon planning for a one product uncapacitated master scheduling problem under stochastic demand. They show that explicitly considering demand stochasticity within the lot sizing model is superior to adding safety stocks in terms of costs.

The case of several items is considered in \cite{Brandimarte.2006}, where a plant-location-based lot sizing formulation for multi-item single-echelon production systems under multi-stage demand uncertainty is proposed. The production plans obtained by solving the stochastic models are compared to the results of a deterministic model. The respective production decisions are applied in a rolling horizon manner. Within the simulation, customer demand is assumed to follow a normal distribution and respectively scenario trees are sampled according to this assumption. Multiple production stages and demand updates, as they are used within our work, are not considered. 

\citet{gansterer_simulation-based_2014} present different approaches to efficiently determine important MRP parameters as they are used in hierarchical production planning, namely planned lead time, safety stock and lot sizes. After analyzing a regular grid over the possible parameter settings in order to find the most suitable setting, they apply a variable neighborhood search procedure to find the best setting in a shorter amount of time. While this work exclusively focuses on planning by means of MRP with well chosen parameters, we additionally solve mathematical lot sizing models and investigate the obtained production plans.

\subsection{Enhancing and Replacing the MRP Logic}
\textcolor{black}{Beyond parameter tuning, a substantial body of literature has addressed MRP's finite capacity limitations through several streams:
\begin{enumerate}[(i)]
    \item Extending MRP logic via capacity checks, alternative routing, dynamic lead times, or load-dependent order release \citep{Taal.1997, Pandey.2000, Kanet.2010, Zapfel.1993, wiendahl1995load, Jodlbauer.2012, Rossi.2017};
    \item Adjusting uncapacitated plans to feasible schedules at bottleneck resources \citep{Tardif.1997, Wuttipornpun.2004, Wuttipornpun.2007, Ornek.2006, Sukkerd.2016};
    \item Hierarchical production planning (HPP), decomposing the problem so that aggregate capacity plans constrain detailed MRP executions \citep{Hax1973HierarchicalIO, Bitran.1993, DauzerePeres.1994, Tremblet.2022, Tremblet.2024};
    \item Advanced inventory management, including multi-echelon safety stock optimization \citep{Meistering.2017, Meistering.2020, TavaghofGigloo.2021} and hybrid push-pull systems such as CONWIP or DDMRP \citep{Spearman.1990, Ptak.2011}.
\end{enumerate}
A further stream replaces MRP entirely with optimization-based models, ranging from LP-based order release \citep{Vo.2003} to multilevel lot-sizing formulations \citep{Billington.1983, Maes.1991, Tempelmeier.1997} and stochastic capacitated lot sizing (SCLSP) using two-stage stochastic programming \citep{Tempelmeier.2007, Helber.2013, Thevenin.2021}. However, most approaches either ignore stochasticity, rely on restrictive assumptions, or are not evaluated in dynamic rolling-horizon simulations with continuous forecast evolution and execution friction. The present work addresses this gap by comparing standard MRP against optimization-based planning within a rolling-horizon discrete-event simulation under forecast-induced demand uncertainty.}

\subsection{Integrated and Simulation-Optimization Approaches}
One example is the integrated lot sizing and scheduling problem, which is studied in \cite{curcio_adaptation_2018}. The authors investigate static robust and two-stage stochastic optimization approaches as approximation strategies to the multi-stage setting. 
The integrated lot sizing and cutting stock problem under demand uncertainty is studied in \cite{curcio_integrated_2022}. The authors apply a robust and two-stage stochastic optimization approach in a rolling horizon manner, in order to adapt the models to the multi-stage stochastic setting.

\citet{quezada_multi-stage_2020} consider multi-stage stochastic programming for an uncapacitated remanufacturing environment. Among other parameters, customer demand is assumed to be uncertain. Scenario trees are used to represent this uncertainty and demands are sampled from a uniform distribution. The authors develop a branch and cut algorithm and apply the proposed multi-stage stochastic model in a rolling horizon manner. Capacity restrictions are not considered, even though they constitute a major limitation in real-world environments.

\citet{simonis_simulationoptimization_2023} investigate simulation-optimization for tablet production in the pharmaceutical industry. In this work, optimization is not applied in a rolling horizon setting. Instead, the authors solve the lot sizing problem for the full planning horizon, simulate the obtained production plan, use the simulation results to update the optimization model and start over the next optimization run. This is done until the optimization result does not change anymore. Customer demand is assumed to be uncertain and a variable neighborhood search procedure to solve the lot sizing problem in a generalized uncertainty framework is proposed. 

The forecast evolution model is considered in the work of \cite{grunow_dynamic_stochastic_2023}. They study the differences of using the additive and multiplicative version of forecast evolution in dynamic lot sizing under demand uncertainty. The authors investigate a multi-stage demand uncertainty setting, however modify the model by splitting the time horizon in two parts. A piecewise-linear approximation is used to determine production lots in the short term, whereas a scenario-tree representation is used to model the demand uncertainty for the later periods.

\citet{almeder2009simulation} investigate simulation and optimization in a supply chain network context. They consider several suppliers, production facilities and customers, while modelling the material flow via different transportation modes. Optimization and simulation are applied iteratively, each considering a different level of detail, using mixed integer linear programming (MILP) models to obtain decision rules for a discrete-event simulation environment. With this approach it is possible to treat nonlinearities, complex structures, stochasticity in the simulation environment, while the optimization is able to provide optimal solutions for a simplified version of the problem.  

For a broad overview of the full spectrum of simulation-optimization methods we refer to available survey papers, such as \cite{figueira_hybrid_2014}. The authors explore the characteristics of the different approaches and provide a systematic classification and taxonomy. Also \cite{Juan.2015} review literature that combines optimization techniques with simulation in order to make better decisions in uncertain real world environments. For general literature on the lot sizing problem we refer to the comprehensive surveys of \cite{karimi_capacitated_2003}, \cite{jans_modeling_2008} and \cite{quadt_capacitated_2008}. 

Even though mathematical optimization models for lot sizing problems have already been applied to stochastic demand situations in a rolling horizon setting, to the best of our knowledge, we are the first to investigate different customer update behaviours and system utilizations. Moreover, we evaluate the proposed planning strategies not only by drawing random demand realizations, but by using discrete-event simulation. 

\section{Simulation-optimization framework}
\label{sec:simulation-optimization_framework}

This section details the elements of the created simulation model, starting with demand generation using forecast evolution.

\subsection{Forecast evolution model}
\label{forecastEvolutionModel}
\textcolor{black}{To capture the effects of forecast evolution on medium-term production planning, customer demand is modeled as a stochastic, time-dependent process. Instead of static inputs, demand info evolves over the horizon to analyze how forecast updates impact planning stability, generated via the additive Martingale Model of Forecast Evolution (MMFE) \citep{Heath.1994,Norouzi.2014}.}

For the applied MMFE the variable $D_{itb}$ in Equation \eqref{eq:demandForecast} denotes the demand forecast for item $i$ for due date $t$, which is provided $b$ periods before delivery. The variable $F_{it}$ represents the long-term forecast of end item $i$ for due date $t$. \textcolor{black}{The long-term forecast} is a constant model parameter for end items, and set in a way that it equals the expected order amount $E[D_{it0}]$. The forecast update horizon, denoted by $T$, is the period during which forecast updates are performed. For periods beyond the forecast horizon $t > T$, customers provide the long-term forecast $F_{it}$\textcolor{black}{, which in the MMFE context represents the initial demand expectation that remains unchanged due to the absence of updated demand information. } Within the applied simulation model for $T$ the constant value of \( {{\demandInformationHorizon}}\) periods is set. 
 
\begin{equation}
{D_{itb}} = \left\{ 
\begin{array}{ll}
{F_{it}} & \text{for } b > T \\
{{F_{it}} + {\varepsilon _{itb}}} & \text{for } b = T \\
{D_{it,b + 1} + {\varepsilon _{itb}}} & \text{for } b < T 
\end{array}
\right.
\label{eq:demandForecast}
\end{equation}
The random update term applied during $b \leq T$ is denoted as $\varepsilon_{itb}$ in Equation \eqref{eq:epsilon_distribution}. \SP{It} is modelled as a truncated normally distributed random variable with mean $0$ and a standard deviation $\sigma_{\varepsilon}$, which is defined as the $\alpha$ fraction of the long-term forecast $F_{it}$. In this setting, $\alpha$ represents the level of unsystematic forecast error. The unsystematic forecast error is the unpredictable, random deviation in forecasting outcomes \citep{Zeiml.2019,Altendorfer.Felberbauer.2023}.
Note that for the forecast updates $\varepsilon_{itb}$, a truncated normal distribution is necessary to avoid negative demand forecast values, i.e., $\varepsilon_{itb} > (-D_{it,b+1})$ and $\varepsilon_{itb} < D_{it,b+1}$. Setting $\varepsilon_{it0}=0$ indicates the final order amount, with no more updates in the delivery period. The expected order amount $E[D_{it0}]$ is set to the corresponding demand behavior quantities for the end items in the numerical study.
\begin{equation}
{\varepsilon _{itb}} \sim N\left( {0,{\sigma _\varepsilon }} \right);{\rm{ }}{\sigma _\varepsilon } = \alpha {F_{it}}
\label{eq:epsilon_distribution}
\end{equation}

\SP{We test} different values for $\alpha$ in combination with three different customer behaviors. More detailed information on the investigated customer behaviours can be found in Section \ref{sec:customer_types_and_demand_patterns}. 

\subsection{Material Requirements Planning (MRP)}

\textcolor{black}{Medium-term planning, as part of hierarchical production planning, addresses decisions over a horizon of several weeks to months \citep{Luo.2022}. In this context, MRP provides detailed material replenishment plans to align internal and external supply and demand \citep{DOLGUI2007269}. Its primary objective is to minimize inventory based on lot-sizing policies while explicitly considering lead times and maintaining safety stocks \citep{Hopp.2011, Orlicky.1975, Vollmann.1997}. A major advantage of standard MRP is its algorithmic simplicity, which iteratively generates material plans regardless of system complexity. However, standard MRP has notable limitations. Crucially, it lacks capacity constraints by decoupling fixed planned lead times from time-dependent department workloads, which can result in practically infeasible manufacturing plans \citep{Zijm.1996}. Furthermore, it is highly susceptible to planning nervousness and the bullwhip effect triggered by changes in the Master Production Schedule (MPS) \citep{Rahman.2020}. Based on \cite{Hopp.2011}, our discrete-event simulation implements standard MRP by processing customer demands within the planning horizon through four sequential steps: (i) netting gross requirements, (ii) applying lot-sizing policies (e.g., FOQ or FOP), (iii) time-phasing via backward scheduling with fixed planned lead times, and (iv) BOM explosion. This procedure is executed iteratively, progressing from end items at BOM level 0 down to the final component level.}

\subsection{Optimization models}

Next to standard MRP, we also obtain production plans by formulating and solving MILP models, which we integrate into the developed simulation-optimization framework. Our models are based on the multi-item multi-echelon capacitated lot sizing formulations of \cite{Thevenin.2021}. Table \ref{tab:problem_parameters} provides the necessary notation. 

\textcolor{black}{To ensure computational tractability and avoid excessive MILP-solver run times  across 400 optimization runs per replication, we adopt a standard capacitated lot-sizing formulation without setup carry-overs. In our setting of small lots and frequent changeovers, saving one boundary setup yields negligible capacity gains. To strictly isolate capacity dynamics, monetary setup costs ($s_{ik}$) are zero. Consequently, the objective minimizes only inventory holding and backlog costs, penalizing changeovers solely through bottleneck capacity ($C_{kt}$) consumption by setup times ($t_{ik}$), as enforced by constraint (\ref{deterministic_capacity_contraint}).}

\begin{table}[ht]
\begin{tabular*}{\hsize}{@{\extracolsep{\fill}}ll@{}}
\hline
Index set & Definition\\
\hline
$t, \tau \in \mathcal{H} = \{1,\ldots,T\}$ & Planning horizon \\
$i,j \in \mathcal{I} = \mathcal{I}_e \cup \mathcal{I}_c$ & Items \\
$\mathcal{I}_c = \{1,\ldots,M\}$ & Components \\
$\mathcal{I}_e = \{{M\!+\!1},\ldots,{M\!+\!N}\}$ & End items \\
$k \in \mathcal{K} = \{1,\ldots,J\}$ & Resources \\
$ \mathcal{K}_i$ for $i \in \mathcal{I}$ & Resources that are needed to produce item $i$ \\
$ \mathcal{I}_k$ for $k \in \mathcal{K}$ & Items that are produced on resource $k$ \\
$\omega \in \Omega$ & Set of scenarios \\
\Rev{$\tilde{T} \in \{1,\ldots,T\}$} & \Rev{Number of periods with assumed fixed production orders} \\
\hline
Parameter & Definition \\
\hline
    $D_{it}$  & Demand of end item $i$ in period $t$\\
    $\hat{I}_{i0}$ & Initial inventory of item $i$ \\
    $\hat{I}_{it}$ & Arriving production order of item $i$ in period $t$ \\
    $R_{ij}$ & Number of units of item $i$ needed for 1 unit of $j$ (BOM)\\
    $L_i$ & Lead time of item $i$ \\
    $C_{kt}$ & Capacity of resource $k$ in period $t$ \\
    $s_{ik}$ & Setup cost of item $i$ for resource $k$ \\
    $t_{k}$ & Setup time for resource $k$ \\
    $p_{i}$ & Production time of item $i$  \\
    $v_i$ & Production cost of item $i$ \\
    $h_i$ & Holding cost of item $i$ \\
    $b_i$ & Backlog cost of end item $i$ \\
    $e_i$ & Lost sales cost of end item $i$ (backlog at period $T$) \\
\hline 
\end{tabular*}
\caption{Problem parameters}
\label{tab:problem_parameters}
\end{table}

\textcolor{black}{Both models presented below schedule a set of items $\mathcal{I}$ over a discrete time horizon $\mathcal{H}$ on capacitated resources $\mathcal{K}$ to satisfy uncertain customer demand, while minimizing setup, production, inventory, backlog, and lost sales costs. In optimization, 'lost sales' serve strictly as an end-of-horizon mathematical penalty to prevent artificially delaying production (horizon effect). Conversely, the discrete-event simulation carries backlogs forward with tardiness penalties, evaluating lost sales penalties only for unfulfilled orders remaining at the absolute end of the simulation runtime.}

\textcolor{black}{Limited capacity and production lead times create a fundamental planning trade-off: incurring inventory holding costs via early production versus risking capacity bottlenecks and subsequent tardiness. Unmet demand accumulates backlog costs $b_i$ per period. Any backlog remaining at the horizon's end becomes lost sales, incurring a significantly higher penalty $e_i$. Allowing lost sales ensures mathematical feasibility even with insufficient capacity. Ultimately, the objective is to determine production quantities that respect operational constraints while minimizing total costs.}

\subsubsection{The deterministic optimization model}
\label{section:deterministic_model}

Decisions made in the deterministic model are based on the latest available information update, without considering possible changes in the future. Table \ref{tab:deterministic_model_variables} shows the variables used in the deterministic optimization model, which is formulated in \eqref{deterministic_objective_function} - \eqref{deterministic_domain_backlog}.

\begin{table}[ht]
\begin{tabular*}{\hsize}{@{\extracolsep{\fill}}ll@{}}
\hline 
Variable & Definition \\
\hline
    $Y_{it}$ & Binary setup variable of item $i$ in period $t$ \\
    $Q_{it}$ & Production quantity of item $i$ in period $t$\\
    $I_{it}$ & Inventory of item $i$ in period $t$\\
    $B_{it}$ & Backlog of end item $i$ in period $t$\\
\hline
\end{tabular*}
\caption{Deterministic model variables}
\label{tab:deterministic_model_variables}
\end{table}

\begin{equation}
\label{deterministic_objective_function}
    \min \sum_{i \in \mathcal{I}_e \cup \mathcal{I}_c} \sum_{t \in \mathcal{H}} \left ( \sum_{k \in \mathcal{K}_i}  s_{ik} Y_{it} +
     v_i Q_{it} + h_i I_{it} \right )+ \sum_{i \in \mathcal{I}_e} \left(\sum_{t=1}^{t=T-1} b_i B_{it} + e_i B_{iT}\right)
\end{equation}
such that
\begin{align}
\label{deterministic_end_item_balance}
    \sum_{\tau=1}^{t-L_i}Q_{i\tau} + \hat{I}_{i0} + \sum_{\tau=1}^{t}\hat{I}_{i\tau} - \sum_{\tau=1}^t D_{i\tau} - I_{it} + B_{it} &= 0 &&  i \in \mathcal{I}_e, t \in \mathcal{H} \\
\label{deterministic_component_balance}
    \sum_{\tau=1}^{t-L_i}Q_{i\tau} + \hat{I}_{i0} + \sum_{\tau=1}^{t}\hat{I}_{i\tau} - \sum_{\tau=1}^t \sum_{j \in \mathcal{I}_e \cup \mathcal{I}_c} R_{ij} Q_{j\tau} - I_{it} &= 0 && i \in \mathcal{I}_c, t \in \mathcal{H} \\
\label{deterministic_setup_constraint}
   Q_{it} - M_{it} Y_{it} &\leq 0 && i \in \mathcal{I}, t \in \mathcal{H} \\
\label{deterministic_capacity_contraint}
    \sum_{i \in \mathcal{I}_k} \textcolor{black}{(} \textcolor{black}{t_k} Y_{it} + \textcolor{black}{p_i} Q_{it} \textcolor{black}{)} &\leq C_{kt} && t \in \mathcal{H}, k \in \mathcal{K} \\
\label{deterministic_domain_setup}
    Y_{it} &\in \{0,1\} && i \in \mathcal{I}, t \in \mathcal{H} \\
\label{deterministic_domain_quantity}
    Q_{it} &\geq 0 && i \in \mathcal{I}, t \in \mathcal{H} \\
\label{deterministic_domain_inventory}
    I_{it}&\geq 0 && i \in \mathcal{I}, t \in \mathcal{H} \\
\label{deterministic_domain_backlog}
    B_{it} &\geq 0 && i \in \mathcal{I}_e, t \in \mathcal{H}
\end{align} 

The objective \eqref{deterministic_objective_function} is to minimize the total costs over the planning horizon, which consist of setup, production, inventory, backlog, and lost sales costs. While production and inventory costs incur for all items, backlog costs only incur for end items. Backlog in the last period of the planning horizon is denoted separately as lost sales. Constraints \eqref{deterministic_end_item_balance} and \eqref{deterministic_component_balance} define the inventory balance for end items and components respectively, taking into account lead times. Constraints \eqref{deterministic_setup_constraint} make sure that items can only be produced if the suitable setup is made in the respective period. The capacity constraints \eqref{deterministic_capacity_contraint} restrict the combined consumption of resource capacity by setup time and production time for each period and resource. \textcolor{black}{In alignment with standard MLCLSP assumptions \citep{Billington.1983}, each item corresponds to an operation that is assigned to exactly one specific capacity-constrained resource. Therefore, the capacity constraints ensure that the sum of setup and processing times for all items assigned to resource $k$ does not exceed its available capacity. This structural assumption holds true for the stochastic model presented in Section \ref{section:stochastic_model} as well.} Finally, the variable domains are specified in \eqref{deterministic_domain_setup} - \eqref{deterministic_domain_backlog}.

\subsubsection{The stochastic optimization model}
\label{section:stochastic_model}
 
To effectively consider uncertainty in customer demands, fluctuations need to be anticipated long before the actual due date. Reaction times in production systems with several echelons are usually longer than one period, meaning that a demand update shortly before the actual due date cannot be compensated anymore by updated optimization runs. The used mathematical model should make use of the available information on the parameter's stochasticity at every call. To realize this, a suitable approach is two-stage scenario-based stochastic optimization, see e.g. \cite{birge_louveaux_2011}. It takes into account the stochasticity by sampling discrete scenarios from an assumed underlying probability distribution and incorporating a metric over these scenarios, such as the expected value, into the optimization model. We denote a scenario by $\omega$ and corresponding demand realization as $D_{it}^{\omega}$, which represents the realized demand of end item $i \in \mathcal{I}_e$ in period $t \in \mathcal{H}$ for scenario $\omega$. The set of all scenarios included in the optimization model is denoted as $\Omega$, with the number of scenarios being the cardinality $|\Omega|$ of this set. Within the optimization model each scenario $\omega$ is weighted with a probability of $p_{\omega}$. In the following, we describe a stochastic version of the model presented in Section \ref{section:deterministic_model}, using two-stage scenario-based stochastic programming. In two-stage models, decision variables are split into first stage and second stage decisions. 

\Rev{In the two-stage stochastic programming model of \citet{Thevenin.2021} the setup and quantity decisions form the first stage decisions, while inventory and backlog are put into the second stage. Quantities are therefore assumed to be fixed over the whole planning horizon, while the scenario dependent inventory and backlog variables are used to estimate the costs related to the discrete demand scenarios. \citet{schlenkrich_capacitated_2024} propose a model that assumes flexibility in production quantities and only fix setup decisions in the first stage, allowing to adapt production quantities as soon as demand is known. Production orders at the beginning of the planning horizon are made \emph{here-and-now}, while production quantities of later periods are naturally flexible, because the optimization model will be solved again with updated information at a later point in time. In order to adequately reflect this setting within the two-stage optimization model, we introduce an additional parameter $\tilde{T}$ representing the period up to which production quantities are assumed to be fixed within the model. Instead of fixing the production quantities $Q_{it}$, associated with item $i \in \mathcal{I}$, for all periods $t \in \mathcal{H}$ in the first decision stage, we only put a part of the quantity decisions $Q_{it}$ with $t \leq \tilde{T}$ in the first stage, while moving the decisions for the remaining periods $t \in \{\tilde{T}+1,\ldots, T\}$ to the second stage. It is necessary to put at least the production quantities of the initial period $t=1$ in the first stage, because these decisions are directly translated into production orders within the simulation.}
Table \ref{tab:stochastic_variables} shows the used decision variables for the stochastic model, which is  formulated in \eqref{stochastic_objective_function}-\eqref{stochastic_domain_backlog}.

\textcolor{black}{In rolling horizon planning, replanning periodicity and the frozen interval ($F$) govern the trade-off between stability and reactivity. While longer frozen intervals mitigate system nervousness \citep{Zhao.1993}, their schedule design is critical under capacity constraints and demand uncertainty \citep{Xie.2003}---our core setting. By setting both replanning periodicity and the physical execution interval $F$ to 1, we maximize reactivity but risk operational nervousness. To counteract this, our stochastic optimization employs the parameter $\widetilde{T}$ \citep{schlenkrich_capacitated_2024}. Crucially, while $F$ defines the actual locked execution periods on the shop-floor, $\widetilde{T}$ acts as an internal non-anticipativity horizon, forcing identical, scenario-independent decisions for the first $\widetilde{T}$ periods. This regularization mathematically replicates the stability of a longer frozen interval without sacrificing physical agility ($F=1$). Systematically varying $\widetilde{T}$ thus reveals how artificial planning stabilization secures coordination and prevents nervousness in complex, capacity-constrained multi-item systems.}

\begin{table}[ht]
\begin{tabular*}{\hsize}{@{\extracolsep{\fill}}ll@{}}
\hline 
Variable & Definition \\
\hline
    $Y_{it}$ & Binary setup variable of item $i$ in period $t$ \\
    $Q_{it}^{\Rev{\omega}}$ & Production quantity of item $i$ in period $t$ \Rev{for scenario $\omega$}\\
    $I_{it}^{\omega}$ & Inventory of item $i$ in period $t$ for scenario $\omega$\\
    $B_{it}^{\omega}$ & Backlog of end item $i$ in period $t$ for scenario $\omega$ \\
\hline
\end{tabular*}
\caption{Stochastic model variables}
\label{tab:stochastic_variables}
\end{table}

\begin{equation}
\begin{aligned}
\label{stochastic_objective_function}
    \min &\sum_{i \in \mathcal{I}_e \cup \mathcal{I}_c} \sum_{t \in \mathcal{H}} \sum_{k \in \mathcal{K}_i} s_{ik} Y_{it} \\
    &+ \sum_{\omega \in \Omega} p_{\omega} \left (\sum_{i \in \mathcal{I}_e \cup \mathcal{I}_c} \sum_{t \in \mathcal{H}} (v_i Q_{it}^{\omega} + h_i I_{it}^{\omega}) + \sum_{i \in \mathcal{I}_e} \left( \sum_{t=1}^{t=T-1} b_i B_{it}^{\omega} + e_i B_{iT}^{\omega} \right) \right)
\end{aligned}
\end{equation}
such that
\begin{align}
\label{stochastic_end_item_balance}
    \Rev{\sum_{\tau=1}^{t-L_i}Q_{i\tau}^{\omega} + \hat{I}_{i0} + \sum_{\tau=1}^{t}\hat{I}_{i\tau} - \sum_{\tau=1}^t D_{i\tau}^{\omega} - I_{it}^{\omega} + B_{it}^{ \omega}} & \Rev{= 0 }&& \Rev{i \in \mathcal{I}_e, t \in \mathcal{H}, {\omega \in \Omega}} \\
\label{stochastic_component_balance}
    \Rev{\sum_{\tau=1}^{t-L_i}Q_{i\tau}^{\omega} + \hat{I}_{i0} + \sum_{\tau=1}^{t}\hat{I}_{i\tau} - \sum_{\tau=1}^t \sum_{j \in \mathcal{I}_e \cup \mathcal{I}_c} R_{ij} Q_{j\tau}^{\omega} - I_{it}^{ \omega}} &\Rev{= 0 }&& \Rev{i \in \mathcal{I}_c, t \in \mathcal{H}, { \omega \in \Omega} }\\
\label{stochastic_non_anticipativity}
    \Rev{Q_{it}^{\omega} }&\Rev{= Q_{it}^{\omega'} }&& \Rev{i \in \mathcal{I}, t \leq \tilde{T}, { \omega, \omega' \in \Omega}}\\
\label{stochastic_setup_constraint}
    \Rev{Q_{it}^{\omega} - M_{it} Y_{it}} &\Rev{\leq 0 }&& \Rev{i \in \mathcal{I}, t \in \mathcal{H}, \omega \in \Omega }\\
\label{stochastic_capacity_contraint}
    \Rev{\sum_{i \in \mathcal{I}_k} \textcolor{black}{(} \textcolor{black}{t_k} Y_{it} + \textcolor{black}{p_i} Q_{it}^{\omega} \textcolor{black}{)} }&\Rev{\leq C_{kt} }&&\Rev{ t \in \mathcal{H}, k \in \mathcal{K}, \omega \in \Omega}\\
\label{stochastic_capacity_contraint}
    \Rev{\sum_{i \in \mathcal{I}_k} \textcolor{black}{(} \textcolor{black}{t_k} Y_{it} + \textcolor{black}{p_i} Q_{it}^{\omega} \textcolor{black}{)} }&\Rev{\leq C_{kt} }&&\Rev{ t \in \mathcal{H}, k \in \mathcal{K}, \omega \in \Omega}\\
\label{stochastic_domain_setup}
    Y_{it} &\in \{0,1\} && i \in \mathcal{I}, t \in \mathcal{H} \\
\label{stochastic_domain_quantity}
Q_{it}^{\omega} &\geq 0 && i \in \mathcal{I}, t \in \mathcal{H}, \omega \in \Omega \\
\label{stochastic_domain_inventory}
I_{it}^{\omega} &\geq 0 && i \in \mathcal{I}, t \in \mathcal{H}, \omega \in \Omega \\
\label{stochastic_domain_backlog}
B_{it}^{\omega} &\geq 0 && i \in \mathcal{I}_e, t \in \mathcal{H}, \omega \in \Omega
\end{align} 
The objective \eqref{stochastic_objective_function} of the stochastic optimization model is to minimize the total resulting costs over the planning horizon consisting of setup costs, as well as the expected \Rev{production}, inventory, backlog and lost sales costs over all scenarios $\omega \in \Omega$. The inventory balance constraints for end items and components need to hold for all scenarios $\omega \in \Omega$ and are transformed to \eqref{stochastic_end_item_balance} and \eqref{stochastic_component_balance}. We note that production quantity variables, as well as inventory and backlog variables are equipped with a scenario index $\omega \in \Omega$. In order to guarantee that quantity variables for periods up to $\tilde{T}$ are actually implementable here and now, we introduce non-anticipativity constraints \eqref{stochastic_non_anticipativity} to force quantity variables $Q_{it}^{\omega}$ with $t \leq \tilde{T}$ to have the same value for all scenarios $\omega, \omega' \in \Omega$. These constraints make the corresponding decision variables equivalent to first stage variables. Setup and resource capacity constraints need to hold for each scenario $\omega \in \Omega$ and are transformed to \eqref{stochastic_setup_constraint} and \eqref{stochastic_capacity_contraint}. While variable domains for setups \eqref{stochastic_domain_setup} remain unchanged from the deterministic model, the domains for quantity, inventory and backlog variables are modified to \eqref{stochastic_domain_quantity}, \eqref{stochastic_domain_inventory} and \eqref{stochastic_domain_backlog}.

To reduce the size of the stochastic model, in our implementation we follow \cite{Thevenin.2021} by implicitly modelling the non-anticipativity constraints. For each item $i \in \mathcal{I}$ and $t \leq \tilde{T}$ we replace the set of quantity variables \{$Q_{it}^{\omega} | \omega \in \Omega\}$ by the single variable $Q_{it}$. By doing so, the non-anticipativity constraints are implicitly considered in the design of the decision variables.

\label{section:simulation_model_and_interaction}
\subsection{Simulation-optimization framework}

\textcolor{black}{Discrete-event simulation (DES) is a well-established method for evaluating production systems \citep{Altendorfer.2016,Altendorfer.Felberbauer.2023,Enns.2001,Juan.2015}, effectively modeling long-term behavior under planning approaches like MRP \citep{Seiringer.Algorithm,Werth.2023}, RPS \citep{Seiringer.WSC23}, CONWIP \citep{Xanthopoulos.2021}, or MILP \citep{gansterer_simulation-based_2014}. Integrating DES with optimization in a rolling horizon requires seamless bidirectional data exchange: the simulation feeds real-time states (demand updates, inventory, WIP, and capacity utilization) to the optimizer, which in turn returns actionable production orders. To maximize shop-floor reactivity, the physical execution frozen interval is set to $F=1$, releasing planned orders immediately at time $t$. The resulting risk of system nervousness is counteracted internally via the stochastic non-anticipativity parameter $\widetilde{T}$, successfully decoupling physical agility from planning stability. Once released, each order is processed lot-for-lot. Its total duration, $\mathit{Dur} = t_k + (Q_{it} p_i)$ (with $t_k$ as stochastic setup time), must elapse entirely before the lot transfers to the next stage. This ensures the simulation accurately reflects the timing assumptions of the optimization models. Appendix \ref{app:planning_interaction} details these technical interaction mechanics.}

\subsubsection{Simulation-optimization component interaction}
\begin{figure}
    \includegraphics[width=\textwidth]{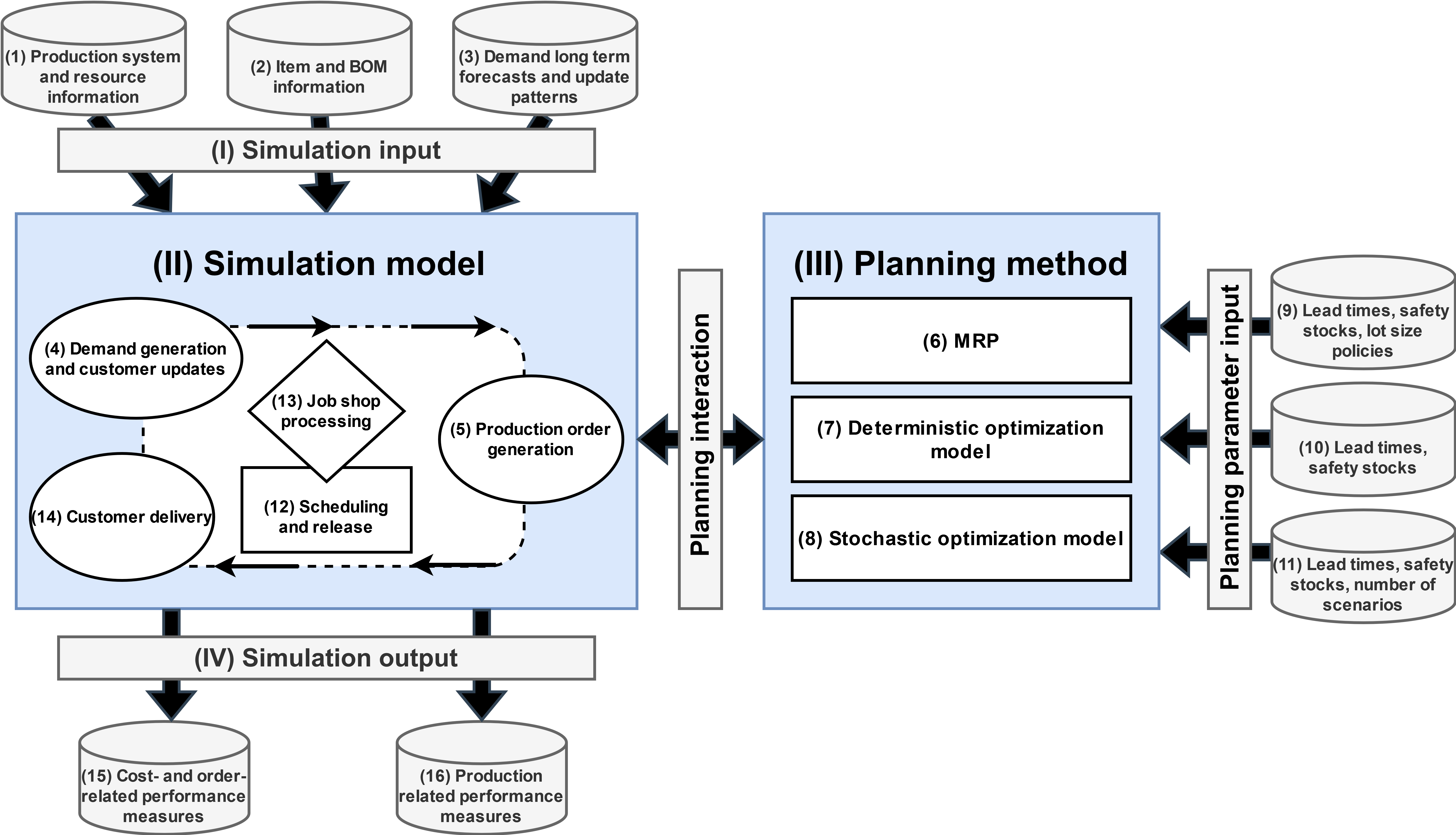}
    \caption{Component interaction of simulation-optimization framework}
    \label{fig:sim-opt-framework}
\end{figure}

The simulation model is provided in form of a discrete-event simulation model developed with Anylogic. In Figure \ref{fig:sim-opt-framework} a schematic representation of the  simulation-optimization framework implemented in Anylogic is shown.

\textcolor{black}{The use of DES is essential to capture the intra-period dynamics that a period-based 'bucket' simulation overlooks. Specifically, DES enables the modeling of stochastic setup times and granular sequencing rules (e.g., Earliest Due Date) at the machine level. This level of detail is necessary to account for boundary effects, where stochastic delays push lot completion across period boundaries, thereby directly impacting inventory levels and tardiness costs.}
This is also the order in which the components are interacting. First the simulation input is provided to the simulation model which iteratively calls the planning method and finally stores the simulation output in a relational database, after the simulation run is finished. The simulation input comprises three types of input data: 
\begin{enumerate}
\item Production system information concerning available resources and their respective capacities.
\item BOM and routing information, specifying which resources are required to processes the items.
\item Customer demand information, including long-term forecasts \textcolor{black}{per end item}, as well as frequency and level of demand updates. 
\end{enumerate}

\textcolor{black}{Once these inputs are established, the \textit{simulation model (II)} runs until period $n$, continuously executing four main steps:}

\begin{enumerate}
    \item \textcolor{black}{Demand Generation (4): Creates customer orders based on the forecast evolution (Section \ref{forecastEvolutionModel}).}
    \item \textcolor{black}{Order Generation (5): Uses a selected \textit{planning method (III)} to create production orders. Options are \textit{MRP (6)} in AnyLogic, or \textit{deterministic (7)} / \textit{stochastic optimization (8)} via Python/CPLEX.}
    \item \textcolor{black}{Scheduling \& Processing (12, 13): Orders are sorted by Earliest Due Date (EDD), released to machine queues, and processed in the job shop.}
    \item \textcolor{black}{Delivery (14): Finished items are stocked for customer delivery, and service levels are measured against due dates.}
\end{enumerate}

\textcolor{black}{Finally, the simulation outputs key performance indicators (KPIs) for costs, orders (e.g., inventory, tardiness), and production (e.g., machine utilization).}

\section{\textcolor{black}{Baseline simulation study setup}}
\label{sec:simiulationStudySetup}
Each replication of the simulation ran for $n=400$ periods (equivalent to weeks), with the first 40 periods serving as a warm-up phase, i.e., $z=40$. \textcolor{black}{To maintain computational feasibility across the simulation runs, we applied a 2\% MIP gap for the stochastic model, and a 5\% MIP gap alongside a 5-second time limit for the deterministic model. Given its lower complexity, the deterministic model typically solves to optimality well before this time limit is reached.}

\textcolor{black}{The simulation accounts for two sources of uncertainty: customer demand forecasts and log-normally distributed setup times. The log-normal distribution is chosen for its positive skewness and zero lower bound, making it highly suitable for modeling production variables \citep{Limpert_Log_normal_distribution}.} \textcolor{black}{We introduce stochastic setup times as "execution friction" to prevent the simulation from perfectly mirroring the optimization models. This model-to-reality gap tests baseline resilience to operational noise and allows us to demonstrate (Section \ref{sec:Results for the mixed production structure}) how a probabilistic feedback loop ensures robust plans under combined uncertainty.} The processing time of items and components, is assumed to be deterministic, assuming a stable production process. To capture the stochastic effects, 10 replications were used for each iteration.

\subsection{Production system}
\label{sec:ProdSys1}
\begin{figure}[ht]
\begin{center}
\includegraphics[scale=0.10]{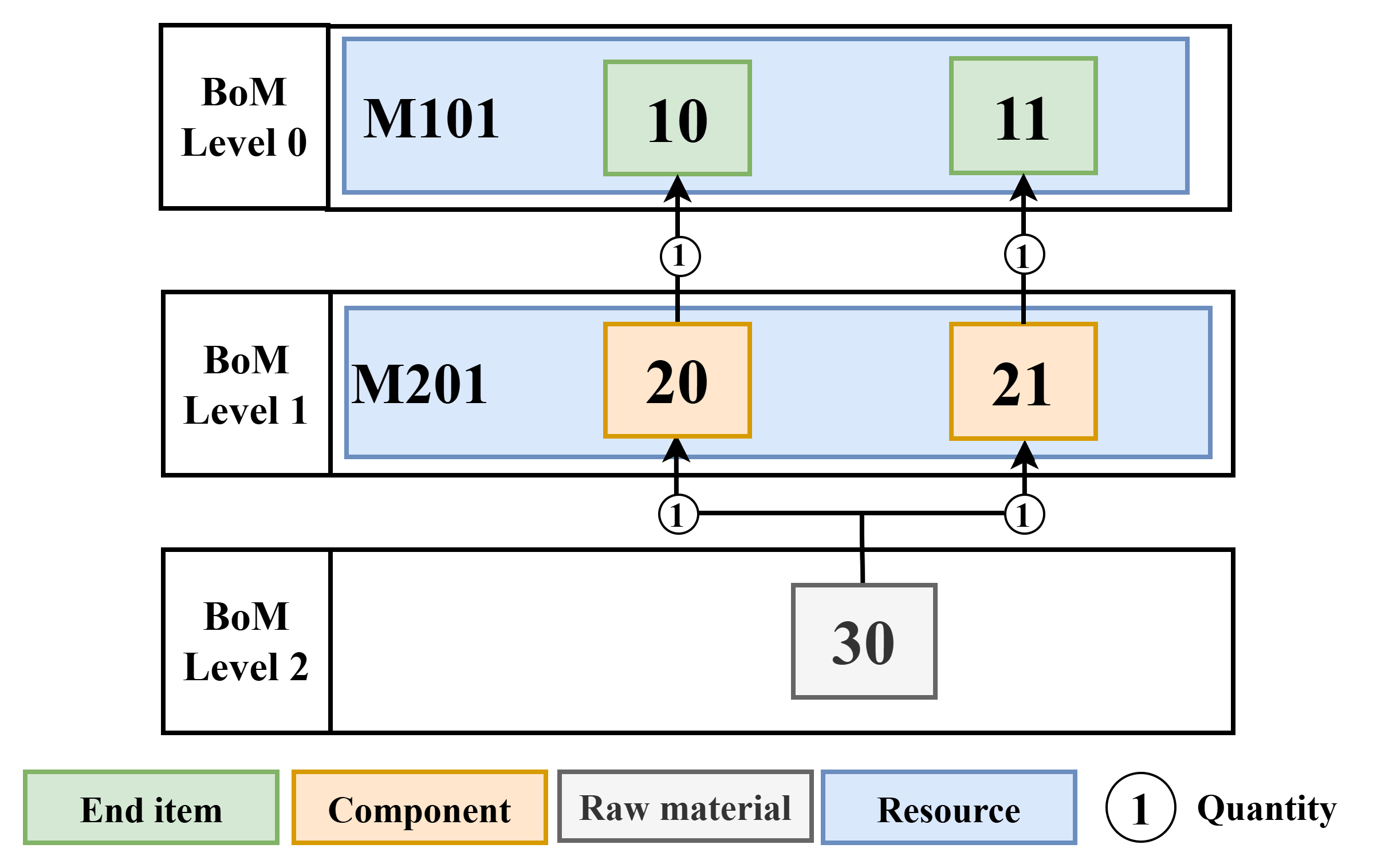}
\caption{\textcolor{black}{BOM and routing information for simple divergent 3-level BOM production system }}
\label{fig:BOM}
\end{center}
\end{figure}

The elementary multi-item and multi-stage
production system is illustrated in Figure~\ref{fig:BOM}. It is important to recognize that smaller systems provide a clearer understanding of how planning parameter settings, influence outcomes \citep{Vidal.2022}. The end items $10$ and $11$, at BOM level $0$ are produced on machine \emph{M2\textcolor{black}{01}}. The components $20$ and $21$ are produced on machine \emph{M1\textcolor{black}{01}}. For one unit of end item $10$, we need $1$ unit of component $20$. The raw material \textcolor{black}{30} is assumed to be always available.

For the production system, a base demand scenario has been established, in which customers consistently order on average 200 units of item 10 and 400 units of item 11 each period (week). This baseline demand scenario is diversified by incorporating various customer behaviors, as detailed in Section \ref{sec:customer_types_and_demand_patterns}. To compute evaluated planned shop loads of 85\%, 90\%, 95\% and 98\% for the demand scenario, the following settings are used: end item and component processing times in minutes per shop load are 1.56 minutes for 85\%, 1.68 minutes for 90\%, 1.8 minutes for 95\% and  1.872 minutes for 98\%. For all items a setup time of 144 minutes is applied. The value of 144 is based on a period capacity of 1440 minutes and a setup proportion of 10\%. To parameterize the applied log-normal distribution for setup time, the coefficient of variation (CV) is set to 20\%, resulting in a standard deviation of 28.8 minutes. This makes the production system capable of producing 600 unites per period (week). 
These settings reflect the aimed production system behavior of one setup in every planning period. \textcolor{black}{Because setup costs are not applied ($s_{ik} = 0$), the optimization models prioritize minimizing inventory levels. This leads to a natural lot-for-lot production behavior under stable, deterministic conditions. This configuration ensures that the models' performance is evaluated based on their ability to navigate capacity bottlenecks and demand volatility, rather than traditional cost-based lot consolidation.}

\subsection{Customer types and demand parameters}
\label{sec:customer_types_and_demand_patterns}

In our numerical experiment, we study three customer behaviours. \textcolor{black}{These three customer behaviors  are selected to reflect common industrial planning situations ranging from stable contractual demand (Type A) to late demand clarification (Type B) and continuously evolving order information (Type C), which are frequently observed in medium-term planning practice.} Customer Type A represents a very reliable customer that only once changes the long-term forecast with the beginning of the demand information horizon when $T$ is 12. This mimics a deterministic demand setting, since there are no demand updates within the planning horizon. 
\textcolor{black}{Customer Type A provides a deterministic baseline and control group with stable demand to verify framework consistency. Under perfect information, optimization results should align with standard lot-for-lot approaches (MRP with FOP 1), validating the mathematical model before introducing the stochastic complexities of Types B and C.}
Customer Type B can only give a rough estimation of the required amount in advance at the beginning of the demand information horizon, when $T$ is 12. The actual demands at the due dates can vary significantly from the first submitted demand estimation, which is modelled by a second demand update right before the respective due date.  Finally, Customer Type C changes demand information frequently and updates the required amount in every period of the planning horizon. 
Furthermore, we study different demand variation factors $\alpha$. Variation factors represent the percentage of the long-term forecast by which the demand changes at every demand update step. The higher these $\alpha$ values are, the more demand uncertainty has to be absorbed by the production system. We vary $\alpha$ within the range \{0.025, 0.05, 0.075, 0.1, 0.125\}. 

\subsection{Planning parameters}
In addition to customer types and demand variation factors, planning parameters for MRP and parameters for both deterministic and stochastic optimization are evaluated. Parameter ranges for this evaluation were chosen based on preliminary simulations.

For MRP, the evaluated parameters include safety stock (SS), lot-sizing policies---fixed order period (FOP) and fixed order quantity (FOQ)---and planned lead time (LT). The ranges for SS are tested as multiples of 0, 0.1, 0.2, 0.3, 0.4, 0.5, and 0.6 of the end item's long-term forecast. FOP values range from 1 to 5 periods, and LT from 1 to 4 periods. FOQ is tested at multiples of 0.5, 1.0, 1.25, 1.5, 1.75, and 2 times the long-term forecast. These parameter values are applied consistently across both end items and their associated BOM components.

\textcolor{black}{While LT and SS are constant inputs per simulation run for controlled benchmarking, their actual utilization during execution is dynamic. Specifically, the stochastic optimization model dynamically generates and depletes safety buffers based on demand scenarios (Section \ref{sec:numericalResults}). Furthermore, optimization models treat lead times $L_i$ merely as capacity-feasible offsets with solver-determined start dates, contrasting with MRP's rigidly fixed planned lead times. Because complex queueing dynamics under tight finite capacities cause theoretical safety stock calculations (e.g., guaranteed service approaches) to fail, we determined MRP safety stocks empirically.}

\textcolor{black}{An extensive grid search over more than 504,000 parameter combinations identified the best-performing lot size, LT, and SS configurations for each demand and shop-load scenario, ensuring that the MRP benchmark is calibrated to its best observed performance within the tested parameter space. 
Results confirm expected production logistics behavior: stable Customer Type A performs best with lot-for-lot planning (LT=1, SS=0, FOP=1), whereas volatile Types B and C require increased safety stocks or longer lead times to absorb demand fluctuations at higher utilizations. Full baseline configurations and cost curves are provided in Appendix \ref{app:mrp_basline}.}

For the optimization approaches, testing 10, 20, 30, and 50 scenarios leads to 1,260 parameterizations for the deterministic model and 5,040 for the stochastic model. With each parameterization simulated across 10 replications, the study encompasses a total of 567,000 individual simulation runs (MRP: 504,000; deterministic: 12,600; stochastic: 50,400).

\section{\textcolor{black}{Numerical results for the baseline system}}
\label{sec:numericalResults}
The discussed results represent the overall costs in cost units (CU) per period computed by the sum of finished goods inventory (FGI), Work in Progress (WIP) and tardiness per unit. For end items a FGI costing factor of {{\holdingCosts}} is applied, for the associated WIP a value of {{\holdingCostsWIP}}. The inventory holding costs for components are {{\holdingCostsComponents}} and for components WIP they are {{\holdingCostsWIPComponents}}. Tardiness costs for the end items is set to 38. The relation of end item WIP and end item tardiness costs represents a target service level of 95\%. A service level of 0.95 means there is a 95\% chance that the inventory on hand will satisfy customer needs \cite{Axsater.2015}. The inventory costs of end items and components are twice the WIP value, as it is more costly to store end items or components.

\textcolor{black}{These values were also set for the optimization models and were used to perform the post simulation run cost evaluation. To interpret these costs correctly, it is essential to distinguish between the buffer mechanisms used. In classic MRP logic, safety stock is a fixed parameter that the system continuously tries to maintain. In contrast, our optimization models do not enforce such a minimum-inventory constraint ($I_{it} \geq \mathit{SS}$). Instead, the "safety stock'' serves only as an initial physical buffer. This allows the stochastic model to demonstrate its dynamic autonomy: rather than rigidly replacing stock, it creates "implicit safety buffers'' by producing preemptively based on scenario probabilities to mitigate bottlenecks. This fundamental difference in buffer management is explicitly analyzed in the use-case evaluation of the mixed production system (Section \ref{sec:use_case}.)}
\textcolor{black}{This section proceeds by establishing an MRP baseline for the simple divergent system (Figure \ref{fig:BOM}). We then analyze the stochastic model's sensitivity (scenario count, $\widetilde{T}$) and the impacts of congestion and forecast uncertainty, concluding with a complex use case to evaluate scalability.}

\subsection{Exploring suitable \Rev{parameters for the} stochastic optimization model}

\Rev{\subsubsection{Number of scenarios}
\label{sec:number_of_scenarios}}
A relevant parameter for the stochastic optimization model, introduced in Section \ref{section:stochastic_model}, is the number of demand scenarios included in the model. Within this section we aim at finding a reasonable number of scenarios to include in our stochastic model and perform experiments with the number of scenarios in \Rev{\{10,20,30,50\}.} \Rev{A second important parameter for the stochastic optimization model is the number of periods with fixed, i.e. first stage, production quantities $\tilde{T}$, which is examined in more detailed in Section \ref{sec:number_of_fixed_periods}. In the experiments conducted to determine a suitable number of scenarios we include the extreme cases $\tilde{T}=1$ and $\tilde{T}=12$, which represent the worst and best case in terms of model size respectively. More flexibility, i.e. $\tilde{T}=1$, leads to more second stage decision variables and a larger model, while less flexibility, i.e. $\tilde{T}=12$, results in a smaller model.}

\begin{figure}[ht]
    \centering
    \includegraphics[width=\textwidth]{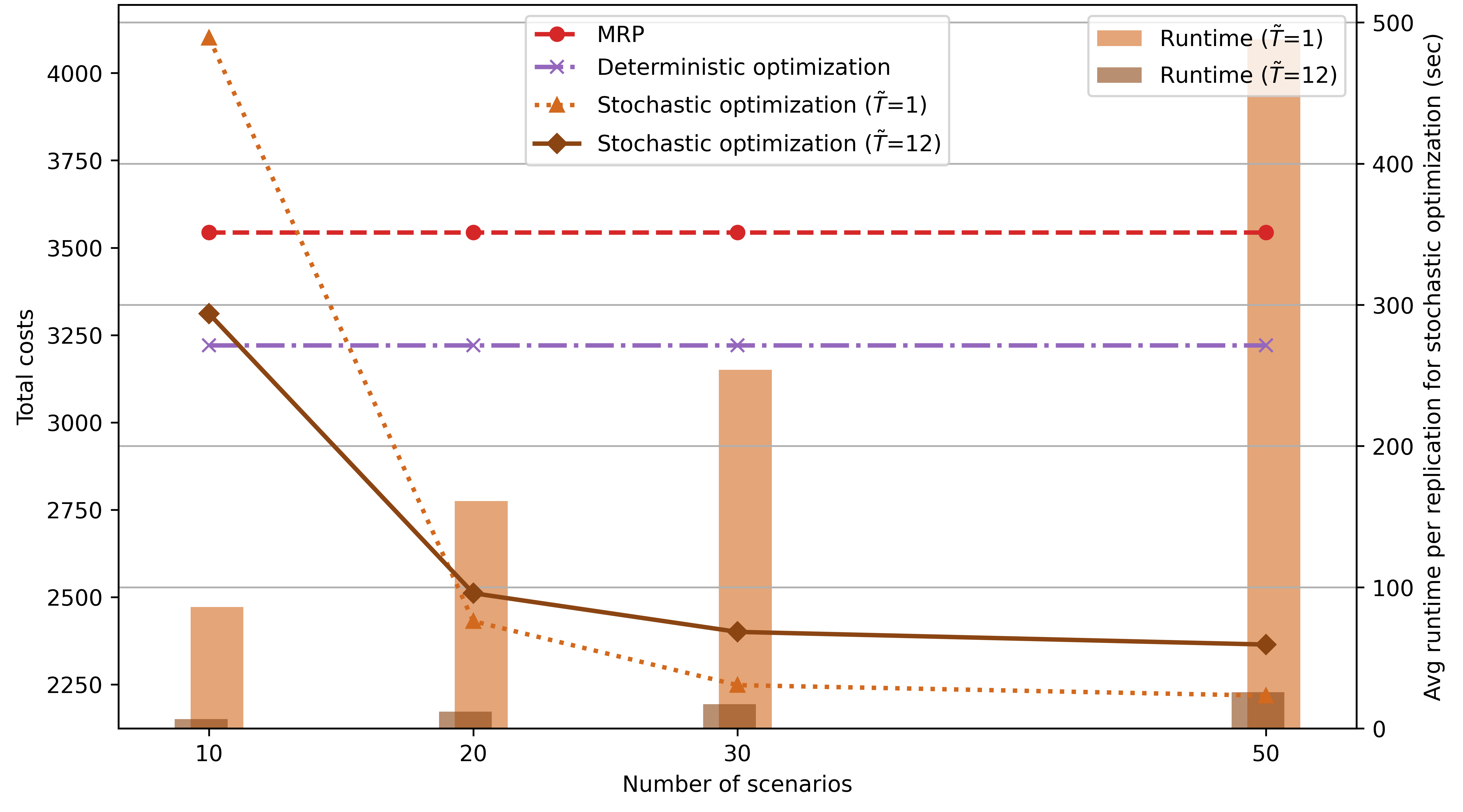}
    \caption{Number of scenarios evaluation for customer update behaviour C: cost and runtime comparison for 95\% utilization and medium demand variation of 7.5\%}
    \label{fig:number_of_scenarios_evaluation_customer_pattern_C}
\end{figure}

We evaluate the case of nervous customer update behaviour C with a medium demand variation of 7.5\% in a production environment with 95\% utilization. Figure \ref{fig:number_of_scenarios_evaluation_customer_pattern_C} shows the obtained total costs of the stochastic optimization model\Rev{s with $\tilde{T}=1$ and $\tilde{T}=12$} for an increasing number of included scenarios. Despite the fact that the performance of the deterministic model, as well as MRP are independent of the number of included scenarios, we report their performance in the investigated setting to put the stochastic performance into perspective. Additionally, on a second y-axis, the average runtime of the stochastic optimization model\Rev{s} is reported for one replication using the respective number of included scenarios. \Rev{Since the stochastic optimization model with $\tilde{T}=1$ is larger than the model with $\tilde{T}=12$ the increase in computational time is an expected result.} We see a strong improvement in total costs when including 20 instead of 10 scenarios, for which the stochastic approach\Rev{es} perform worse than the deterministic model. Increasing the number of scenarios to 30 further reduces the costs, however the improvement is significantly smaller. For the model with $\tilde{T}=12$, the runtime has increased from 6 seconds for 10 scenarios to 17 seconds for 30 scenarios. In the case of the more flexible model with $\tilde{T}=1$, the runtime has increased from 86 seconds for 10 scenarios to 254 seconds for 30 scenarios. While the inclusion of further scenarios only results in slight cost reductions it significantly increases the average runtime for solving the stochastic models, especially in the case of $\tilde{T}=1$. \textcolor{black}{This diminishing return in solution quality relative to the computational effort aligns with established Sample Average Approximation (SAA) literature, which demonstrates that a moderate number of scenarios is generally sufficient to achieve stable solutions in stochastic discrete optimization and lot sizing \citep{Kleywegt.2002}, thereby supporting the structural generalizability of our observations and answers RQ2.}

For this reason we have decided to include 30 scenarios in all following stochastic models, which serves as a reasonable trade-off between solution quality and runtime.

\Rev{
\subsubsection{Number of periods with fixed production quantities}
\label{sec:number_of_fixed_periods}

After determining a suitable number of scenarios for the stochastic optimization model we examine the number of periods $\tilde{T}$ up to which production quantities are assumed to be fixed, i.e. first stage decisions. We repeat the experiments conducted in Section \ref{sec:number_of_scenarios} for utilizations of 85\%, 90\%, 95\% and 98\% with a fixed number of scenarios of 30 and test $\tilde{T} \in \{1,2,\ldots,12\}$. Figure \ref{fig:T-tilde_evaluation_customer_C} shows the respective results. The stochastic model with the highest level of flexibility, i.e. $\tilde{T}=1$, results in the lowest total costs across all utilizations. Especially for higher utilizations also the model that assumes fixed production quantities for the whole planning horizon, i.e. $\tilde{T}=12$, leads to comparably low costs. Fixing a part of the quantity decisions, i.e. $\tilde{T} \in \{2,\ldots, 11\}$ was less successful for the investigated settings.

\begin{figure}[ht]
    \centering
    \includegraphics[width=\linewidth]{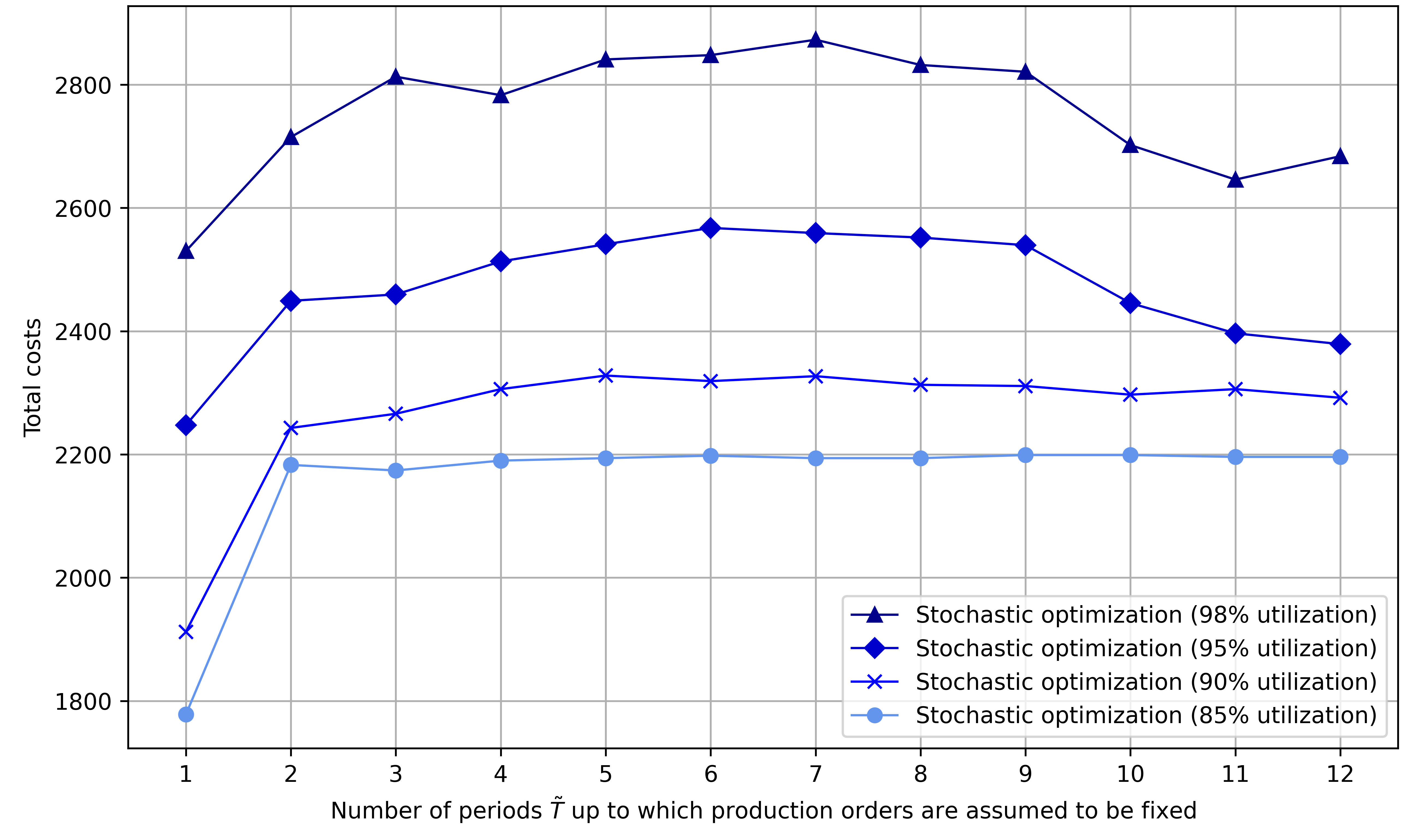}
    \caption{Evaluation of the stochastic optimization model (30 scenarios) for different $\tilde{T}$ values (number of periods with fixed production quantities) without safety stocks: cost comparison for customer update behaviour C and medium demand variation of 7.5\% across utilizations}
    \label{fig:T-tilde_evaluation_customer_C}
\end{figure}

The experiments reported in Figure \ref{fig:T-tilde_evaluation_customer_C} are performed without using safety stocks within the stochastic models. In order to guarantee a fair comparison with MRP, as well as deterministic optimization, we also consider the possibility for safety stocks within the stochastic optimization model. The results are reported in Figure \ref{fig:T-tilde_evaluation_customer_C_inkl_SS} and shows that the presence of safety stocks reduces the effects of assuming flexibility in production quantities. We still identify $\tilde{T}=1$ to be the best choice for resource utilizations of 85\% and 90\%. For a utilization of 95\%, a $\tilde{T}$ value of 4 periods is the best choice, while for the highest utilization of 98\% the choice for 12 periods is most suitable. \textcolor{black}{This confirms that while flexibility is beneficial under lower loads, rising utilization necessitates longer fixation intervals to avoid substantial system nervousness and maintain planning stability. In this context, $\tilde{T}$ acts as an internal stabilizer that replicates the effects of a frozen schedule to ensure feasible capacity coordination and reduce schedule instability under high congestion \citep{Helber.2013, Zhao.2001, Zhao.1997}.} In the remaining part of the computational study we will use the best identified values for $\tilde{T}$, as well as the default value of 12, which is equivalent to fixing quantities of the whole planning horizon. In the case of 98\% resource utilization, where $\tilde{T}=12$ is the best value, we additionally report results for the second best value, which is $\tilde{T}=1$.}

\begin{figure}[ht]
    \centering
    \includegraphics[width=\linewidth]{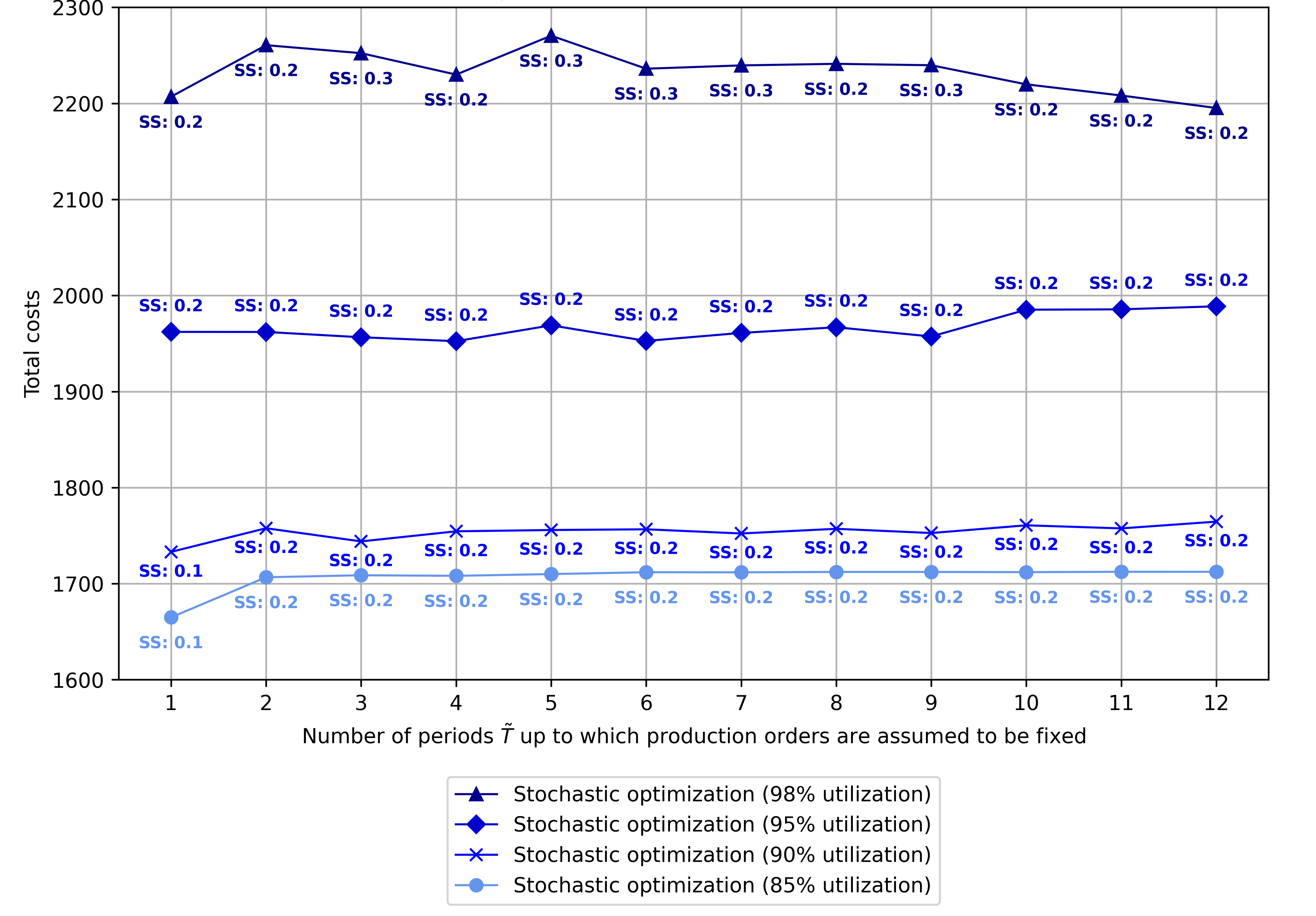}
    \caption{Evaluation of the stochastic optimization model (30 scenarios) for different $\tilde{T}$ values (number of periods with fixed production quantities) considering the possibility of safety stocks: cost comparison for customer update behaviour C and medium demand variation of 7.5\% across utilizations}
    \label{fig:T-tilde_evaluation_customer_C_inkl_SS}
\end{figure}

\subsection{The impact of shop load congestion}
\begin{figure}[ht]
    \centering
        \includegraphics[width=\linewidth]{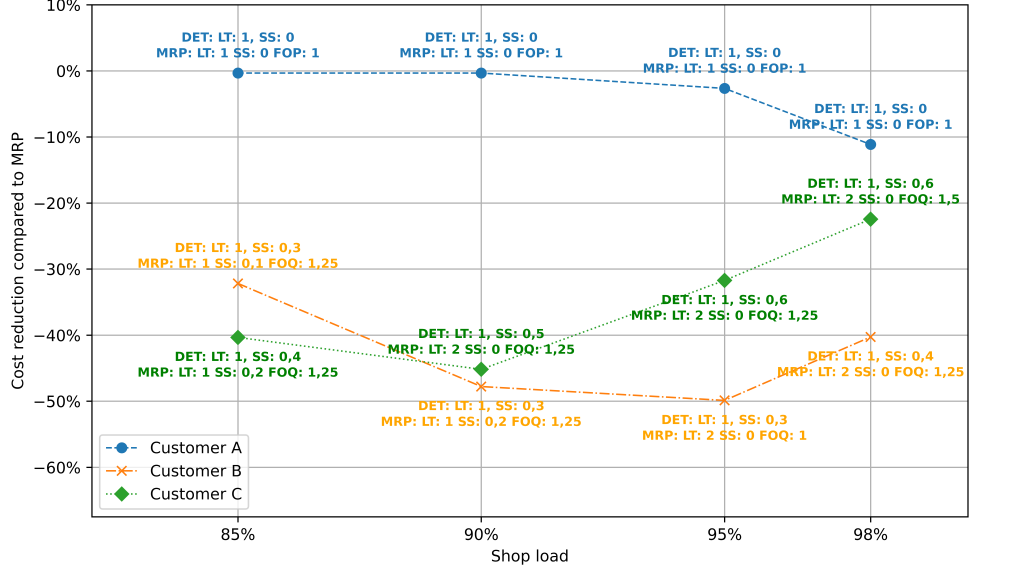}
        \caption{Cost comparison of deterministic optimization to MRP in a setting with medium demand variation (7.5\%) for different shop loads}
        \label{fig:best_performing_deterministic_medium_variation}
        
\end{figure}

\begin{figure}
    \centering
        \includegraphics[width=\linewidth]{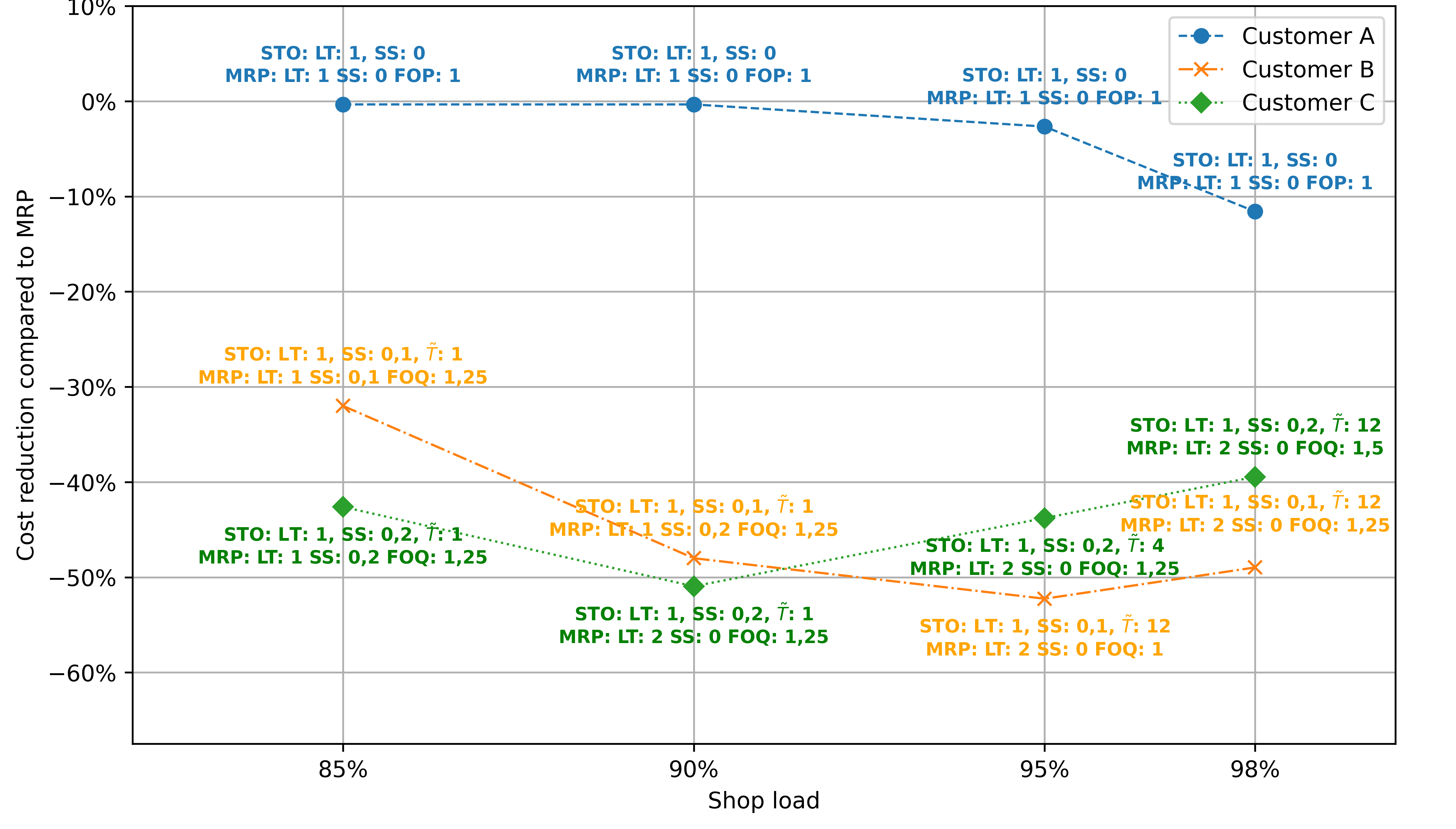} 
        \caption{Cost comparison of stochastic optimization (30 scenarios) to MRP in a setting with medium demand variation (7.5\%) for different shop loads}
        \label{fig:best_performing_stochastic_medium_variation}
\end{figure}

We report the cost reduction of deterministic and stochastic optimization models over standard MRP for Customer Types A, B and C in the case of medium demand variation (7.5\%) and resource utilizations of 85\%, 90\%, 95\% and 98\%. For each of the three planning approaches we use the best performing parameter combination. The results are displayed in Figure \ref{fig:best_performing_deterministic_medium_variation} for the deterministic, while Figure \ref{fig:best_performing_stochastic_medium_variation} shows the results for the stochastic approach. 

We note that the demand pattern of Customer Type A does not include updates, meaning that the stochastic optimization model is equivalent to the deterministic one. For this reason the performance of deterministic and stochastic optimization is equal for Customer Type A, i.e. the same cost reduction is reported in Figure \ref{fig:best_performing_deterministic_medium_variation} and \ref{fig:best_performing_stochastic_medium_variation}. For low resource utilization, the optimization approaches are not able to generate an advantage over standard MRP in the case of Customer Type A. This is reasonable, since customer demand does not necessitate any short-term updates, and the low utilization allows the MRP approach to perfectly fulfill all demands on time.

\textcolor{black}{For Customer Type A, zero setup costs ($s_{ik} = 0$) drive the optimizer to minimize inventory and backlog. Under stable demand, this naturally converges with MRP's lot-for-lot (FOP 1) policy, validating the deterministic model. Meanwhile, stochastic setup times inject operational noise, testing schedule robustness and creating a structural cost floor. As utilization increases, costs rise significantly (Figure 3) due to non-linear queueing effects causing severe WIP and tardiness \citep{Karmarkar.1989, Hopp.2011, Asmundsson.2009, Missbauer.2002}. Because the simulation fulfills all demand---treating backlogs as delays rather than lost sales---minimizing these operational costs remains the sole performance objective.}

Only considering high utilization, we observe limitations of MRP. Using an optimization model allows to consider restrictions and to react to bottleneck situations. For type A, in 98\% utilization, this advantage can reduce costs by 10\% compared to MRP.
 
Customer Type B changes demands last minute. We observe a performance improvement of both optimization approaches compared to MRP. For lower utilizations both models lead to reductions between 30\% and 50\%. The use of safety stocks allows the deterministic model to hedge against demand uncertainty. While the best safety stock for the deterministic model \Rev{lies between 30\% and 40\%}, it is 10\% for the stochastic model. The limitations of the deterministic model become visible in high utilization (98\%). Stochastic optimization is able to outperform MRP by 50\%, the deterministic model results in a reduction of 40\%.

Customer Type C changes demand forecasts every period. Both optimization approaches outperform MRP. The deterministic models leverage safety stocks to hedge against demand variations. While a safety stock of 20\% leads to the best performance of the stochastic model, the deterministic model uses 60\% in high utilization. The performance improvement of the deterministic approach reduces to 32\% for 95\% utilization and 22\% for 98\% utilization. Stochastic optimization leads to a reduction of 40\% in high utilization. \Rev{With rising utilization it is advantageous to increase $\tilde{T}$.} Results for low demand variation (2.5\%) and high demand variation (12.5\%) are reported in Figures \ref{fig:best_performing_deterministic_low_variation} - \ref{fig:best_performing_stochastic_high_variation} in the Appendix. Differentiation between optimization approaches is weaker in the low variation case, it is stronger for high demand variations.

\subsection{The impact of forecast uncertainty}
\begin{figure}[ht]
    \centering
    \includegraphics[width=\textwidth]{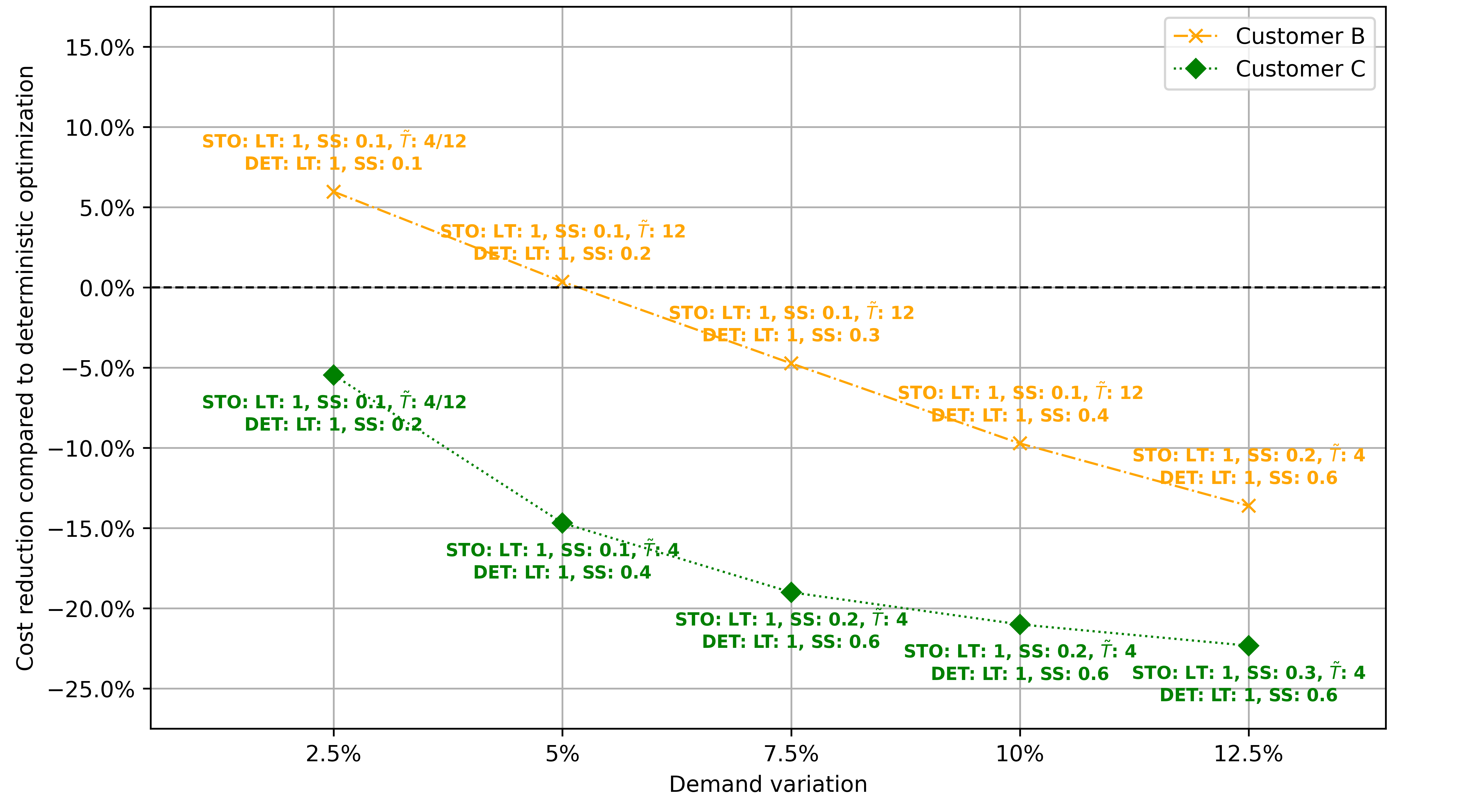}
    \caption{Cost comparison of stochastic optimization to deterministic optimization with 95\% utilization}
    \label{fig:forecast_uncertainty_95}
\end{figure}

To assess the impact of increasing uncertainty in demand forecasts \textcolor{black}{and to answer RQ3}, we perform a series of experiments with demand variation values $\alpha$ in the range \{2.5\%, 5\%, 7.5\%, 10\%, 12.5\%\}. Figure \ref{fig:forecast_uncertainty_95} shows the cost reduction of stochastic optimization over deterministic optimization for utilization of 95\%. Additionally, in Figure \ref{fig:forecast_uncertainty_85} and Figure \ref{fig:forecast_uncertainty_98} we report the results for the low utilization of 85\% and high utilization of 98\% case in the Appendix A.

For Customer Type B deterministic optimization outperforms the stochastic model in low demand variation with 2.5\%. The deterministic model and a safety stock of 10\% leads to a 6\% reduction over the stochastic approach. The deterministic model leverages safety stocks, using 30\% for 7.5\% and 60\% for 12.5\%. This safety stock weakens the deterministic approach and allows the stochastic model, which keeps safety stock at 10\%-20\%, to result in reduced costs. The improvement is 5\% for 7.5\%, it is 14\% with 12.5\%.

Deterministic optimization could not outperform the stochastic approach for Customer Type C in 95\% utilization. For 2.5\%, the stochastic model reduced costs by 5\%. This improvement increased leading to a 23\% reduction for 12.5\%. The deterministic model increases safety stocks, storing 60\% as a buffer. For the stochastic approach, a safety stock of 30\% is sufficient.

\section{\textcolor{black}{Scalability and complex use case evaluation}}
\label{sec:use_case}

\begin{figure}
    \centering
    \includegraphics[width=9cm]{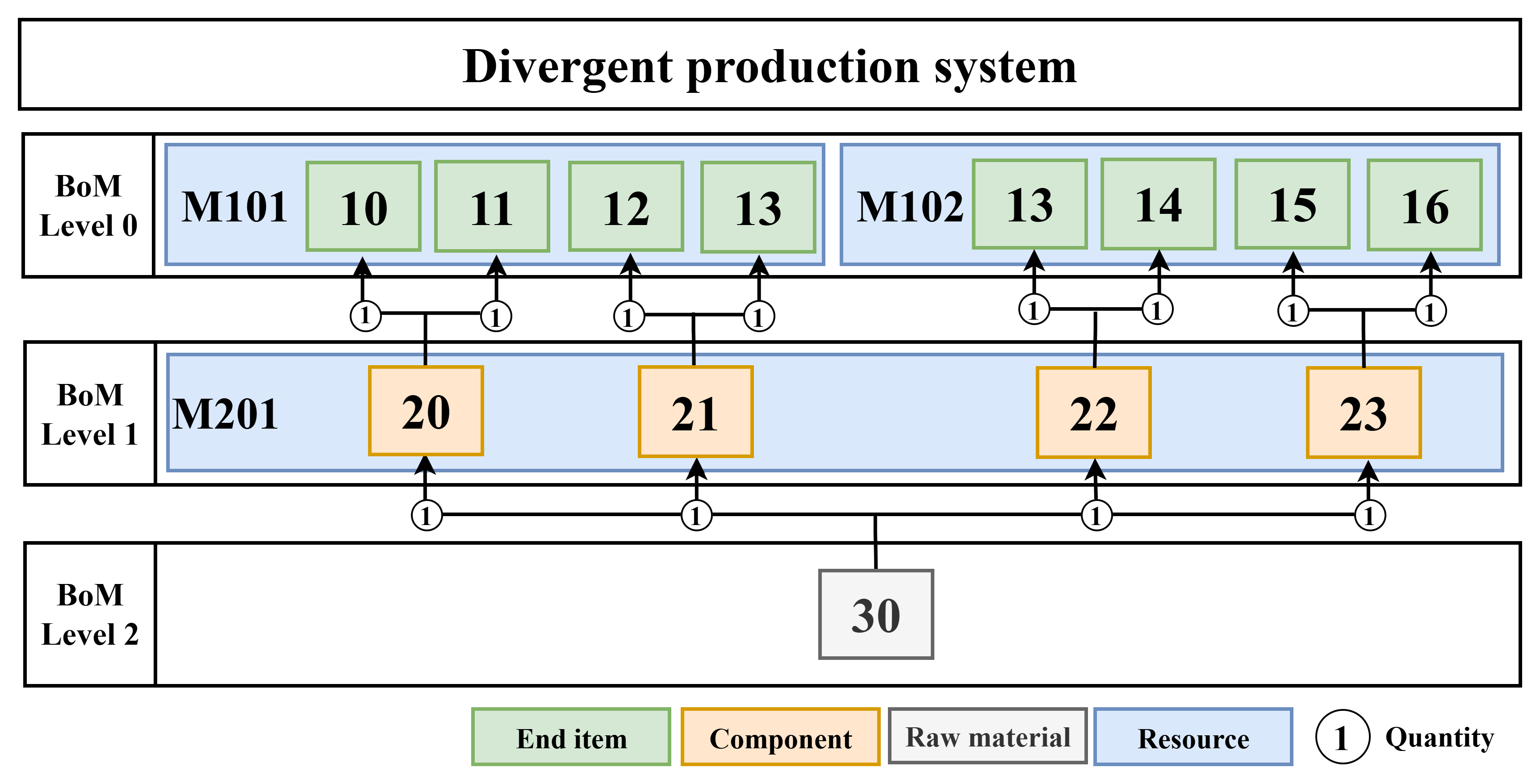}
    \caption{\textcolor{black}{BOM and routing information for divergent production system with 8 end items and 4 components.}}
    \label{fig:medium-scale_production_system}
\end{figure}

\begin{figure}
    \centering
    \includegraphics[width=8cm]{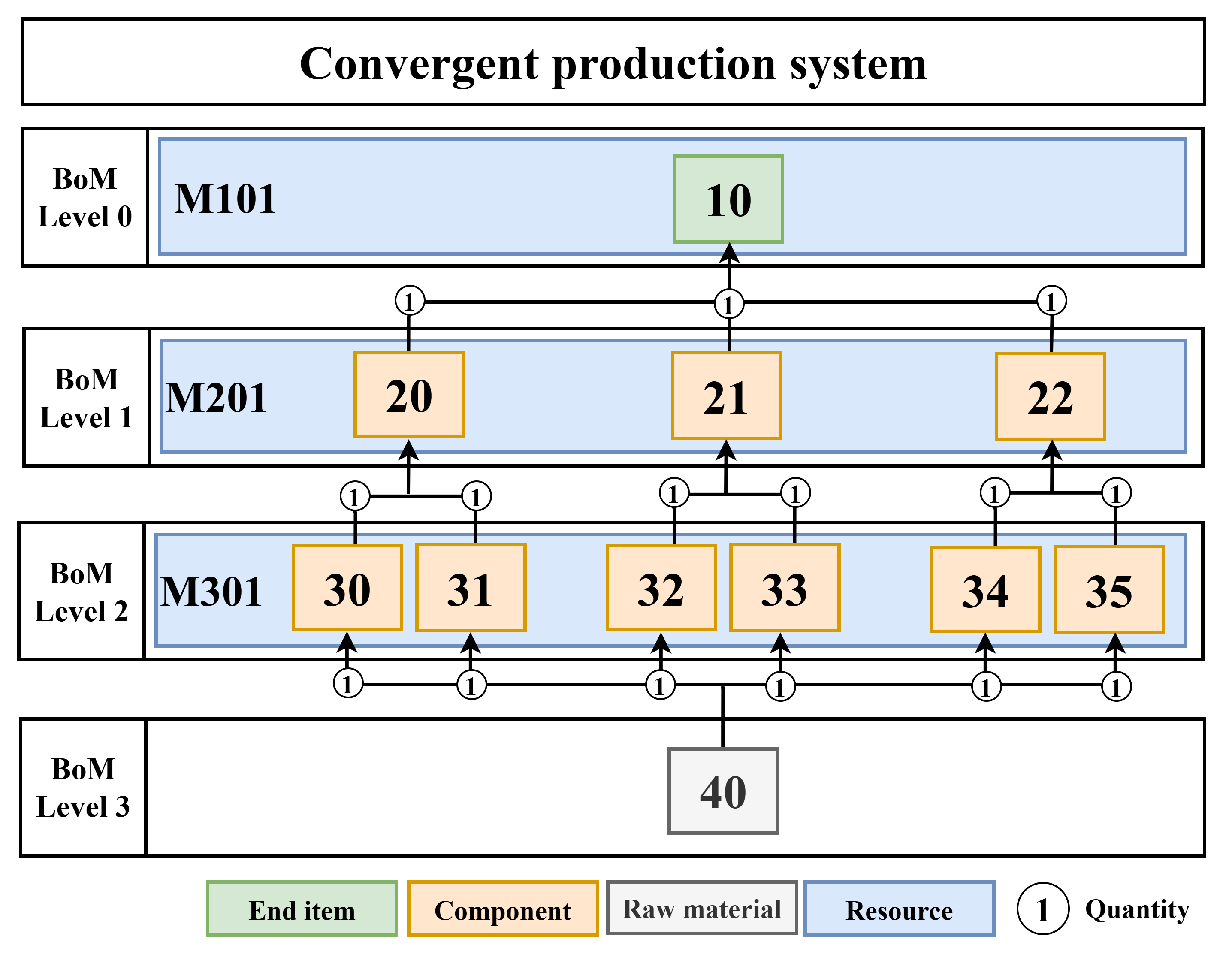}
    \caption{\textcolor{black}{BOM and routing information convergent product structure (assembly type with 1 end item and 9 components.)}}
    \label{fig:convergent_production_system}
\end{figure}

\begin{figure}
    \centering
    \includegraphics[width=8cm]{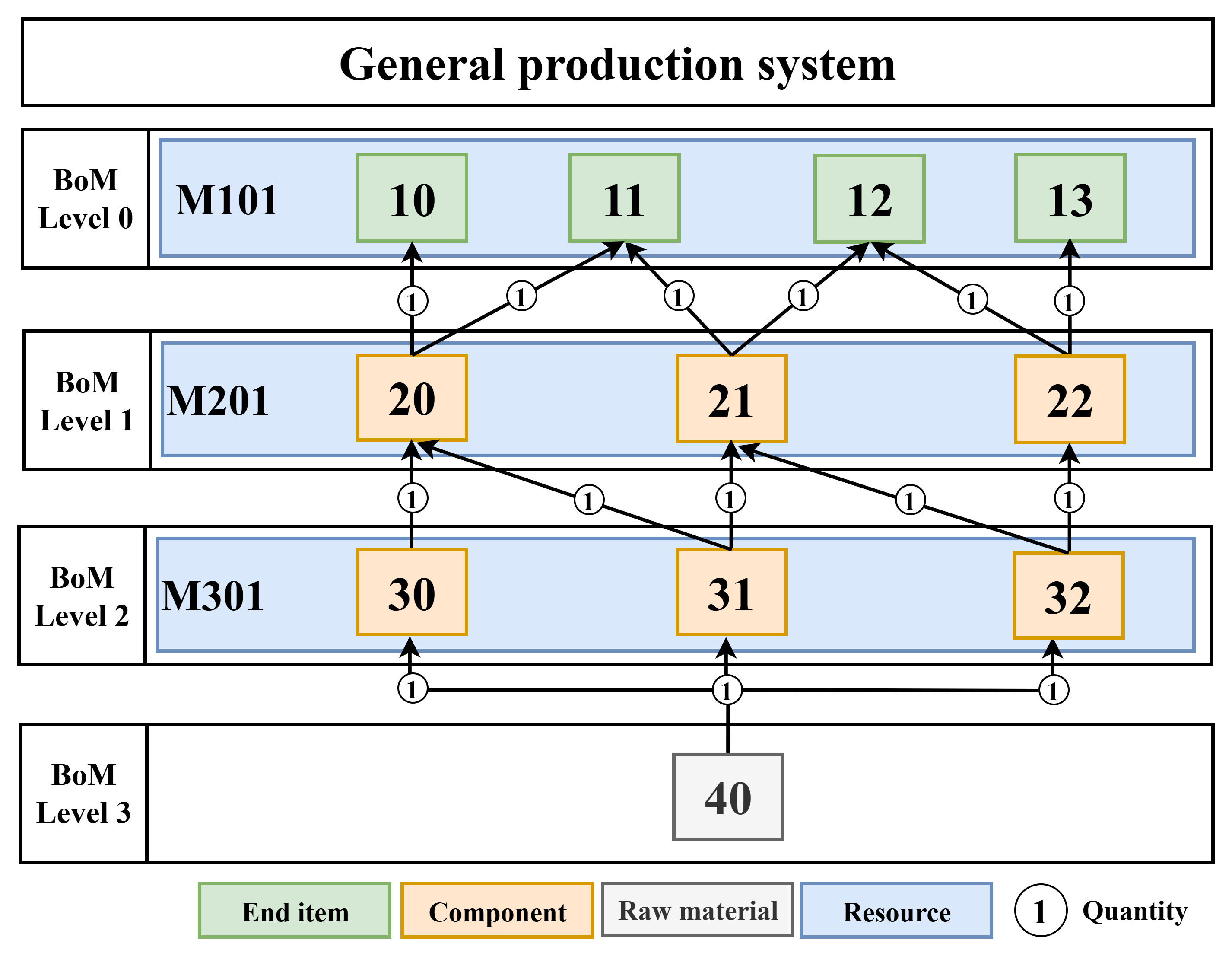}
    \caption{\textcolor{black}{BOM and routing information for general product structure (mixed network type) with 4 end items and 6 components.}}
    \label{fig:mixed_production_system}
\end{figure}

To evaluate the proposed simulation-optimization framework in larger settings, we propose \textcolor{black}{three} additional more complex multi-item multi-stage production systems. \textcolor{black}{First,} Figure \ref{fig:medium-scale_production_system} shows the BOM for \textcolor{black}{a divergent} production system consisting of 8 end items and 4 components produced on 3 different resources. \textcolor{black}{Additionally, Figure \ref{fig:convergent_production_system} illustrates the BOM and routing information for a purely convergent (assembly) production system. It consists of end item 1, produced on resource $M101$, and nine components (20 to 35) produced on resources $M201$ and $M301$.} \textcolor{black}{Furthermore, Figure \ref{fig:mixed_production_system} presents the BOM and routing information for a general production system characterized by a mixture of convergent and divergent elements. It comprises end items 10 to 13 (produced on machine $M101$), components 20 to 22 (produced on machine $M201$), and components 30 to 32 (produced on machine $M301$).}

\clearpage


\textcolor{black}{For the subsequent analysis, we also examine the period $\tilde{T}$ for fixed production quantities in the stochastic optimization model. Unlike the smaller system in Section \ref{sec:ProdSys1} (Figure \ref{fig:BOM}), the three more complex systems (Figures \ref{fig:medium-scale_production_system}, \ref{fig:convergent_production_system}, and \ref{fig:mixed_production_system}) feature varied BOM structures and multiple items sharing resources, so the optimal $\tilde{T}$ may differ among these configurations.} 

In order to evaluate the impact of $\tilde{T}$, we perform experiments with $\tilde{T} \in \{1,\ldots,12\}$ for Customer Type C with medium demand variation of 7.5\%. We report the results for the default value $\tilde{T}$=12, as well as for the best found $\tilde{T}$.  The included number of scenarios in the stochastic model is again 30. For \textcolor{black}{all three} more complex production system\textcolor{black}{s} we test resource utilizations of 80\%, 85\% and 90\%. \textcolor{black}{We cap the utilization at 90\% because our results indicate that the system already exhibits a tardiness share of nearly 50\% at this level, indicating a severe loss of schedule feasibility due to tight capacities. Simulating utilization beyond 90\% in this multi-stage environment degenerates into pure backlog accumulation driven by queueing physics, masking the actual performance differences of the planning methods under evaluation. We report results in Tables \ref{tab:medium-scale_results}, \ref{tab:production_convergent}, and \ref{tab:production_mixed}, providing cost values for MRP, the deterministic, and the stochastic optimization models. For each approach, we display results using the best-found parameterization, both with and without safety stocks, since identifying optimal safety stock levels remains a significant practical challenge in real-world production planning.} 

\subsection{Results for the divergent production system}

\textcolor{black}{Table \ref{tab:medium-scale_results} presents results for the scenario where no safety stocks are applied, providing a baseline to put the findings with safety stocks into perspective.}
We observe that the stochastic optimization model clearly outperforms both, MRP and the deterministic model, in the case of no safety stocks. Applying the deterministic model without safety stock leads to significantly larger costs, also compared to MRP. This is because MRP can compensate the lack of safety stock by adjusting the lot size policy. Applying the stochastic model in this case, leads to cost improvements between 35\% and 54\% compared to MRP. 

\begin{table}[htbp]
    \centering
    \resizebox{\textwidth}{!}{
        \setlength{\tabcolsep}{3pt}
        \begin{tabular}{c | ccccc | cccc | cccc | ccccc | c}
            \hline
            \multicolumn{20}{c}{No safety stocks} \\
            \hline
            & \multicolumn{5}{c|}{MRP} & \multicolumn{4}{c|}{Deterministic} & \multicolumn{4}{c|}{Stochastic ($\tilde{T}=12$)} & \multicolumn{5}{c|}{Stochastic (Best $\tilde{T}$)} & \% $\Delta$ \\
            & Cost & LT & SS & LSP & SL & Cost & LT & SS & SL & Cost & LT & SS & SL & Cost & LT & SS & $\tilde{T}$ & SL & \\
            \hline
            80\% & 21,334 & 2 & 0 & FOQ 2   & 95.62\% & 31,772 & 1 & 0 & 44.83\% & \textbf{9,627} & 1 & 0 & 89.80\% & \textbf{9,627} & 1 & 0 & 12 & 89.80\% & -54\% \\
            85\% & 14,210 & 1 & 0 & FOQ 1.5 & 82.68\% & 30,698 & 1 & 0 & 44.58\% & 8,780          & 1 & 0 & 91.31\% & \textbf{8,767} & 1 & 0 & 6  & 91.24\% & -38\% \\
            90\% & 15,436 & 2 & 0 & FOQ 1.5 & 98.40\% & 32,020 & 1 & 0 & 45.07\% & 10,719         & 1 & 0 & 87.02\% & \textbf{9,931} & 1 & 0 & 10 & 86.68\% & -35\% \\
            \hline
            \multicolumn{20}{c}{With safety stocks} \\
            \hline
            & \multicolumn{5}{c|}{MRP} & \multicolumn{4}{c|}{Deterministic} & \multicolumn{4}{c|}{Stochastic ($\tilde{T}=12$)} & \multicolumn{5}{c|}{Stochastic (Best $\tilde{T}$)} & \% $\Delta$ \\
            & Cost & LT & SS & LSP & SL & Cost & LT & SS & SL & Cost & LT & SS & SL & Cost & LT & SS & $\tilde{T}$ & SL & \\
            \hline
            80\% & 21,334 & 2 & 0   & FOQ 2 & 95.62\% & \textbf{7,409} & 1 & 0.4 & 99.10\% & 7,852 & 1 & 0.15 & 92.89\% & 7,691          & 1 & 0.15 & 1  & 93.10\% & -65\% \\
            85\% & 10,958 & 1 & 0.2 & FOP 2 & 91.33\% & 7,813          & 1 & 0.4 & 97.12\% & 7,531 & 1 & 0.1  & 94.27\% & \textbf{7,519} & 1 & 0.1  & 7  & 94.24\% & -31\% \\
            90\% & 11,958 & 1 & 0.3 & FOP 2 & 89.46\% & \textbf{8,044} & 1 & 0.4 & 95.11\% & 9,275 & 1 & 0.1  & 89.68\% & 8,562          & 1 & 0.1  & 10 & 89.41\% & -32\% \\
            \hline
        \end{tabular}
    }
    \caption{Cost, parameter and service level (SL) overview for the \textcolor{black}{divergent} production system. Comparison of MRP (LT, SS, LSP), deterministic optimization (LT, SS) and stochastic optimization (LT, SS, $\tilde{T}$) (for 30 scenarios) with and without safety stocks.}
    \label{tab:medium-scale_results}
\end{table}

The deterministic model benefits significantly from considering safety stocks.  When choosing a sufficient safety stock level, the performance of the deterministic model is similar to the stochastic optimization model and can lead to cost reductions of up to 65\% compared to standard MRP. It is even able to outperform the stochastic model by around 6\% in certain cases. The stochastic model, however, relies on significantly smaller safety stock levels, by dynamically creating safety buffers to counter demand variations. 

\textcolor{black}{A limited analysis under 80\% utilization with additional SS factors (0.05, 0.125, 0.15, 0.175) shows that 0.15 yields minimum costs for $\tilde{T}=12$, confirming that marginal SS changes significantly alter the stochastic optimization's internal buffer behavior. However, scaling this fine-grained search to all settings would expand parameter steps from 7 to 19 per approach, substantially increasing computational complexity. We therefore retained the original parametrization, leaving a full grid search or more sophisticated tuning for future research and referring to the detailed SS analysis presented for the mixed production system.} When it comes to the most suitable choice of $\tilde{T}$ we observe the same outcome as for the simple divergent production system (Figure \ref{fig:BOM}). For low resource utilization ($80\%$) it is beneficial to keep production quantities flexible ($\tilde{T}=1$), while for increasing utilization it is more suitable to consider fixed production quantities in the majority of the periods ($\tilde{T}=7$ for $85\%$ and $\tilde{T}=10$ for $90\%$). 

\textcolor{black}{These findings are well-supported by existing literature: the 
superiority of stochastic over deterministic planning under uncertainty 
\citep{Brandimarte.2006}, the compensating role of well-calibrated safety stocks 
in multi-stage systems \citep{Inderfurth.1998}, the dynamic buffering capability 
of stochastic lot-sizing models \citep{Helber.2013}, and the need for longer 
frozen planning periods under higher resource utilization \citep{Zhao.2001}.}

\subsection{Results for the convergent production structure}

\textcolor{black}{Table \ref{tab:production_convergent} presents costs and parameters 
for the convergent production system, comparing MRP, deterministic, and stochastic 
optimization with and without safety stocks (baseline without safety stocks reported 
first, as in Table \ref{tab:medium-scale_results}). Without safety stocks, the 
stochastic model again outperforms both alternatives through dynamic buffering 
against demand variability. With safety stocks, the deterministic model improves 
markedly, but the stochastic model continues to excel with lower safety stock 
levels. Unlike the divergent system, $\tilde{T}$ has significantly less influence: 
$\tilde{T}=1$ remains optimal across all settings. This is because the convergent 
system produces only one end item, allowing production adjustments each period 
without resource conflicts. In the eight-item divergent system, low $\tilde{T}$ 
causes planning nervousness, as multiple items competing for shared resources 
induce coordination instability \citep{Kadipasaoglu.1995}. Thus, the convergent 
system benefits from maximum flexibility while the divergent system requires longer 
frozen periods to stabilize multi-product coordination. Overall, $\Delta$ values 
range from -62\% to -68\%, confirming the effectiveness of optimization-based 
planning.}

\begin{table}[htbp]
    \centering
    \resizebox{\textwidth}{!}{
        \setlength{\tabcolsep}{3pt}
        \begin{tabular}{c | ccccc | cccc | cccc | ccccc | c}
            \hline
            \multicolumn{20}{c}{No safety stocks} \\
            \hline
            & \multicolumn{5}{c|}{MRP} & \multicolumn{4}{c|}{Deterministic} & \multicolumn{4}{c|}{Stochastic ($\tilde{T}=12$)} & \multicolumn{5}{c|}{Stochastic (Best $\tilde{T}$)} & \% $\Delta$ \\
            & Cost & LT & SS & LSP & SL & Cost & LT & SS & SL & Cost & LT & SS & SL & Cost & LT & SS & $\tilde{T}$ & SL & \\
            \hline
            80\% & 71,638 & 4 & 0 & FOQ 1 & 99.71\% & 60,462 & 1 & 0 & 42.24\% & 24,579 & 1 & 0 & 87.07\% & \textbf{23,664} & 1 & 0 & 1 & 90.23\% & -67\% \\
            85\% & 71,093 & 4 & 0 & FOQ 1 & 99.71\% & 63,342 & 1 & 0 & 37.93\% & 25,750 & 1 & 0 & 73.28\% & \textbf{24,084} & 1 & 0 & 1 & 74.43\% & -66\% \\
            90\% & 70,341 & 4 & 0 & FOQ 1 & 99.14\% & 68,800 & 1 & 0 & 28.74\% & 29,498 & 1 & 0 & 54.02\% & \textbf{26,782} & 1 & 0 & 1 & 57.18\% & -62\% \\
            \hline
            \multicolumn{20}{c}{With safety stocks} \\
            \hline
            & \multicolumn{5}{c|}{MRP} & \multicolumn{4}{c|}{Deterministic} & \multicolumn{4}{c|}{Stochastic ($\tilde{T}=12$)} & \multicolumn{5}{c|}{Stochastic (Best $\tilde{T}$)} & \% $\Delta$ \\
            & Cost & LT & SS & LSP & SL & Cost & LT & SS & SL & Cost & LT & SS & SL & Cost & LT & SS & $\tilde{T}$ & SL & \\
            \hline
            80\% & 60,096 & 2 & 0.3 & FOQ 1 & 97.99\% & 22,507 & 1 & 0.5 & 88.79\% & 22,360 & 1 & 0.3 & 89.94\% & \textbf{22,135} & 1 & 0.2 & 1 & 92.82\% & -63\% \\
            85\% & 71,093 & 4 & 0.0 & FOQ 1 & 99.71\% & 23,255 & 1 & 0.6 & 77.01\% & 22,580 & 1 & 0.3 & 76.15\% & \textbf{22,438} & 1 & 0.2 & 1 & 76.72\% & -68\% \\
            90\% & 70,341 & 4 & 0.0 & FOQ 1 & 99.14\% & 26,653 & 1 & 0.6 & 56.03\% & 24,084 & 1 & 0.6 & 68.68\% & \textbf{23,935} & 1 & 0.4 & 1 & 66.09\% & -66\% \\
            \hline
        \end{tabular}
    }
    \caption{Cost, parameter and service level (SL) overview for the convergent production system. Comparison of MRP (LT, SS, LSP), deterministic optimization (LT, SS) and stochastic optimization (LT, SS, $\tilde{T}$) (for 30 scenarios) with and without safety stocks}
    \label{tab:production_convergent}
\end{table}

\subsection{Results for the mixed production structure}
\label{sec:Results for the mixed production structure}

\textcolor{black}{Table \ref{tab:production_mixed} reveals patterns largely consistent with the divergent (Table \ref{tab:medium-scale_results}) and convergent systems (Table \ref{tab:production_convergent}). Without safety stocks, stochastic optimization performs best at 80\% and 85\% utilization. However, at 90\% utilization, MRP emerges as superior --- a deviation from earlier trends highlighting 
the impact of tight resource constraints on planning performance \citep{Tempelmeier.2011}. Deterministic optimization without safety stocks clearly underperforms both alternatives. With safety stocks, the competitive landscape changes: deterministic optimization benefits substantially more from their introduction than MRP or stochastic approaches, transforming from suboptimal to 
highly competitive with pronounced cost reductions (high $\Delta$ values) relative to MRP. The stochastic model remains generally robust under uncertainty, but its advantage diminishes under high utilization.}

\begin{table}[htbp]
    \centering
    \resizebox{\textwidth}{!}{
        \setlength{\tabcolsep}{3pt}
        \begin{tabular}{c | ccccc | cccc | cccc | ccccc | c}
            \hline
            \multicolumn{20}{c}{No safety stocks} \\
            \hline
            & \multicolumn{5}{c|}{MRP} & \multicolumn{4}{c|}{Deterministic} & \multicolumn{4}{c|}{Stochastic ($\tilde{T}=12$)} & \multicolumn{5}{c|}{Stochastic (Best $\tilde{T}$)} & \% $\Delta$ \\
            & Cost & LT & SS & LSP & SL & Cost & LT & SS & SL & Cost & LT & SS & SL & Cost & LT & SS & $\tilde{T}$ & SL & \\
            \hline
            80\% & 21,218 & 1 & 0 & FOQ 1.5 & 88.65\% & 37,174 & 1 & 0 & 44.47\% & 16,295 & 1 & 0 & 89.15\% & \textbf{16,295} & 1 & 0 & 12 & 89.15\% & -23\% \\
            85\% & 22,522 & 1 & 0 & FOQ 1.5 & 85.56\% & 38,537 & 1 & 0 & 44.54\% & 18,888 & 1 & 0 & 82.61\% & \textbf{18,819} & 1 & 0 & 3 & 82.97\% & -16\% \\
            90\% & \textbf{25,042} & 1 & 0 & FOQ 1.5 & 82.11\% & 40,232 & 1 & 0 & 45.19\% & 25,811 & 1 & 0 & 65.23\% & 25,502 & 1 & 0 & 7 & 74.43\% & 2\% \\
            \hline
            \multicolumn{20}{c}{With safety stocks} \\
            \hline
            & \multicolumn{5}{c|}{MRP} & \multicolumn{4}{c|}{Deterministic} & \multicolumn{4}{c|}{Stochastic ($\tilde{T}=12$)} & \multicolumn{5}{c|}{Stochastic (Best $\tilde{T}$)} & \% $\Delta$ \\
            & Cost & LT & SS & LSP & SL & Cost & LT & SS & SL & Cost & LT & SS & SL & Cost & LT & SS & $\tilde{T}$ & SL & \\
            \hline
            80\% & 20,969 & 1 & 0.2 & FOQ 1.5 & 93.25\% & \textbf{13,382} & 1 & 0.5 & 99.28\% & 15,259 & 1 & 0.1 & 92.17\% & 15,139 & 1 & 0.1 & 3 & 90.52\% & -36\% \\
            85\% & 22,522 & 1 & 0.0 & FOQ 1.5 & 85.56\% & \textbf{13,243} & 1 & 0.5 & 98.13\% & 17,418 & 1 & 0.2 & 87.07\% & 17,371 & 1 & 0.1 & 3 & 85.56\% & -41\% \\
            90\% & 25,042 & 1 & 0.0 & FOQ 1.5 & 82.11\% & \textbf{13,870} & 1 & 0.5 & 96.41\% & 22,067 & 1 & 0.6 & 74.71\% & 21,825 & 1 & 0.6 & 7 & 83.55\% & -45\% \\
            \hline
        \end{tabular}
    }
    \caption{Cost, parameter and service level (SL)  overview for the mixed production system. Comparison of MRP (LT, SS, LSP), deterministic optimization (LT, SS) and stochastic optimization (LT, SS, $\tilde{T}$) (for 30 scenarios) with and without safety stocks.}
    \label{tab:production_mixed}
\end{table}

\textcolor{black}{We now analyze why stochastic optimization underperforms in highly 
constrained systems, particularly at 90\% utilization where MRP achieves lowest 
costs without safety stocks and deterministic optimization dominates with safety 
stocks. To explain these effects, we examine additional shop-floor KPIs reported 
in Table \ref{tab:mixed_vertical_final_clean}: flow time for end items 
($\mathit{Flow}_{EI}$) and components ($\mathit{Flow}_{Co}$), the synchronization 
gap ($\mathit{SyncGap}$, defined as the difference between these flow times 
capturing temporal misalignment between stages), and lateness (average internal 
end-item delay relative to planned completion). All KPIs are computed after the 
warm-up phase on the first simulation replication. Results show that stochastic 
optimization performance deteriorates with increasing utilization.}

\begin{table}[h]
\resizebox{\textwidth}{!}{
    \centering
    \setlength{\tabcolsep}{2pt}
    \begin{tabular}{c|cc|cc|cc|cc|c}
    \hline
    \multicolumn{10}{c}{Mixed Production System - Performance Overview} \\
    \hline 
    & \multicolumn{2}{c|}{Inventory Metrics} & \multicolumn{2}{c|}{Flow Times} & \multicolumn{5}{c}{Operative KPIs} \\
    & Overall Costs & (Inv\% / Tard\%) & EI & Co & \multicolumn{2}{c|}{SyncGap} & \multicolumn{3}{c}{Lateness} \\
    \hline
    \multicolumn{10}{l}{\textbf{Utilization 80\%}} \\
    \hline
    MRP & 21,218 & (87\% / 13\%) & 0.6460 & 0.6457 & \multicolumn{2}{c|}{0.0003} & \multicolumn{3}{c}{0.0423} \\
    Deterministic & 37,174 & (33\% / 67\%) & 0.5047 & 0.5666 & \multicolumn{2}{c|}{-0.0619} & \multicolumn{3}{c}{0.0000} \\
    Stoch. T=12 & 16,295 & (79\% / 21\%) & 0.5612 & 0.5739 & \multicolumn{2}{c|}{-0.0126} & \multicolumn{3}{c}{0.0216} \\
    Stoch. Best T & \textbf{16,295} & (79\% / 21\%) & 0.5612 & 0.5739 & \multicolumn{2}{c|}{-0.0126} & \multicolumn{3}{c}{0.0216} \\
    \hline
    \multicolumn{10}{l}{\textbf{Utilization 85\%}} \\
    \hline
    MRP & 22,522 & (81\% / 19\%) & 0.7340 & 0.7036 & \multicolumn{2}{c|}{0.0304} & \multicolumn{3}{c}{0.0783} \\
    Deterministic & 38,537 & (31\% / 69\%) & 0.5389 & 0.6062 & \multicolumn{2}{c|}{-0.0673} & \multicolumn{3}{c}{0.0008} \\
    Stoch. T=12 & 18,888 & (68\% / 32\%) & 0.6401 & 0.6160 & \multicolumn{2}{c|}{0.0241} & \multicolumn{3}{c}{0.0510} \\
    Stoch. Best T & \textbf{18,819} & (68\% / 32\%) & 0.6444 & 0.6114 & \multicolumn{2}{c|}{0.0329} & \multicolumn{3}{c}{0.0457} \\
    \hline
    \multicolumn{10}{l}{\textbf{Utilization 90\%}} \\
    \hline
    MRP & \textbf{25,042} & (70\% / 30\%) & 0.8591 & 0.8125 & \multicolumn{2}{c|}{0.0467} & \multicolumn{3}{c}{0.1454} \\
    Deterministic & 40,232 & (30\% / 70\%) & 0.5720 & 0.6532 & \multicolumn{2}{c|}{-0.0812} & \multicolumn{3}{c}{0.0011} \\
    Stoch. T=12 & 25,811 & (51\% / 49\%) & 0.9167 & 0.7568 & \multicolumn{2}{c|}{0.1599} & \multicolumn{3}{c}{0.2430} \\
    Stoch. Best T & 25,502 & (51\% / 49\%) & 0.8054 & 0.8069 & \multicolumn{2}{c|}{-0.0014} & \multicolumn{3}{c}{0.1584} \\
    \hline
    \end{tabular}
    }
    \caption{Performance comparison of the mixed production system across different utilization levels. Overall costs are supplemented by the percentage share of inventory and tardiness costs (Inv\% / Tard\%) within the main metrics. Operative KPIs include Flow Times for End Items (EI) and Components (Co), SyncGap, and Lateness.}
    \label{tab:mixed_vertical_final_clean}
\end{table}

\textcolor{black}{At 90\% utilization, this trend becomes explicit: stochastic optimization exhibits a tardiness share of 49\%, compared to 30\% under MRP, despite comparable inventory shares. This shift is driven by execution-level effects. Under tight capacity conditions, stochastic optimization leads to longer flow times for end items and elevated internal lateness, indicating loss of schedule feasibility. The best stochastic parameterization at 90\% utilization (overall costs = 25,502) shows a slightly negative $\mathit{SyncGap}$ ($-0.0014$), implying that component and end-item flows are nearly aligned on average, with components requiring marginally more time. In contrast, positive $\mathit{SyncGap}$ values observed for other stochastic settings indicate earlier completion of components and congestion in downstream stages. Despite the improved mean synchronization in the best stochastic case, internal lateness remains high ($0.1584$), demonstrating that average flow alignment does not prevent local congestion and delay accumulation under capacity saturation. These results indicate that, under stochastic optimization, scenario-based release decisions interact with stochastic setup times during execution, leading to increased variability in job flow and resource loading that cannot be compensated under the absence of execution-level flexibility. To examine whether this behavior is driven by limited execution-level flexibility, the analysis is extended to explicitly incorporate stochastic setup time variability. \textcolor{black}{This finding is consistent with \citet{Missbauer.2020}, who 
systematically show that planning models degrade under capacity saturation 
when congestion effects from execution-level variability are not explicitly 
integrated, and with \citet{Pahl.2007}, who review extensive evidence that 
lead times increase nonlinearly with utilization, causing planning-level 
decisions to lose effectiveness when shop-floor variability is ignored.}}

\subsubsection{Integrating setup time variability into the optimization model}

\textcolor{black}{To account for setup time variability, realized times from the
simulation are fed back into the optimization model before each planning step.
After the warm-up phase, the ratio between realized and planned setup times is
recorded for each item to estimate the empirical bias $\beta_i$. Assuming a
log-normal distribution with a fixed coefficient of variation (CV) = 0.2
(reflecting moderate industrial variability and ensuring numerical stability),
the adjusted setup time $\tilde{s}_{ik}$ for item $i$ on resource $k$ is:
\begin{equation}
\tilde{s}_{ik} = s_{ik} \cdot \beta_i \cdot e^{(z \cdot \sigma_i - 0.5 \cdot \sigma_i^2)}
\end{equation}
with $\sigma_i = \sqrt{\ln(1 + CV^2)}$. The z-score $z$ controls the
probabilistic protection level against setup time variability and is varied
across $z \in \{0, 0.84, 1.28, 1.64\}$, corresponding to target service
levels of 50\%, 80\%, 90\%, and 95\%, respectively. This effectively transforms
the stochastic capacity requirement into a deterministic equivalent close to robust capacity constraints
with tunable protection levels \citep{Bertsimas.2004}, where $z$ acts
as a safety multiplier on the capacity coefficient. The adjustment is capped at 2.0
to prevent infeasibility and rounded to the nearest integer for solver
compatibility. The iterative update of $\tilde{s}_{ik}$ before each planning
interval synchronizes the optimization model with the stochastic execution
level, following the principle of iterative parameter adaptation in rolling
horizon planning frameworks \citep{Sahin.2013}. By combining empirical bias
correction with quantile-based safety buffers, the framework ensures that
generated schedules remain feasible and robust against realized shop-floor
variability.}

\subsubsection{Overtime enabled capacity constraints}

\textcolor{black}{In production planning, enforcing strict capacity limits under 
uncertainty frequently leads to model infeasibility. To prevent this and 
represent limited short-term capacity flexibility, we extend the base models 
into a Capacitated Lot Sizing Problem with Overtime (CLSP-O) framework 
\citep{Tas.2019}. Specifically, we model capacity limits as soft 
constraints by introducing non-negative slack variables for overtime, denoted 
as $O_{kt}$ (deterministic) and $O_{kts}$ (stochastic). When activated, the 
original strict capacity constraints \eqref{deterministic_capacity_contraint} 
and \eqref{stochastic_capacity_contraint} are relaxed:
\begin{equation}
\sum_{i \in \mathcal{I}} (a_{ik} \cdot X_{it} + s_{ik} \cdot \delta_{it}) \leq C_{kt} + O_{kt}
\end{equation}
\begin{equation}
\sum_{i \in \mathcal{I}} (a_{ik} \cdot X_{its} + s_{ik} \cdot \delta_{its}) \leq C_{kt} + O_{kts}
\end{equation}
To prevent arbitrary use of this flexibility, overtime is penalized linearly 
using a penalty method. The objective functions \eqref{deterministic_objective_function} 
and \eqref{stochastic_objective_function} are extended with the following cost terms:
\begin{equation}
\sum_{t \in \mathcal{T}} \sum_{k \in \mathcal{K}} c^{OT} \cdot O_{kt}
\end{equation}
\begin{equation}
\frac{1}{|\mathcal{S}|} \sum_{s \in \mathcal{S}} \sum_{t \in \mathcal{T}} \sum_{k \in \mathcal{K}} c^{OT} \cdot O_{kts}
\end{equation}
where $c^{OT}$ denotes the overtime cost coefficient. In the context of 
two-stage stochastic programming, $O_{kts}$ functions as a complete second-stage 
recourse decision \citep{birge_louveaux_2011, Tas.2019}. This formulation allows the 
model to adapt flexibly to specific scenario realizations (wait-and-see) 
without rendering the first-stage setup decisions infeasible. The simulation 
tests a spectrum of $c^{OT}$ values ranging from highly accessible (0.5), to 
moderate (7.0, 25.0), up to prohibitive (1000.0), which mathematically enforces 
near-strict capacity limits. When deactivated, $O_{kt}$ and $O_{kts}$ are 
omitted entirely, strictly enforcing the nominal capacity $C_{kt}$.}

\subsubsection{Impact of stochastic setup times and overtime}
\textcolor{black}{To evaluate the combined impact of both adaptations, a full-factorial 
experiment was conducted based on the best-identified parameterizations for the 
mixed production structure (Table \ref{tab:mixed_vertical_final_clean}). Where the optimal $\tilde{T}$ was 12, the 
second-best value was also tested to assess the impact of flexible periods. Results 
are presented first without, then with safety stock.}

\paragraph{Without‑safety‑stock setting}
\textcolor{black}{As shown in Table \ref{tab:safety_stock_comparison_extended}, introducing execution-level adaptations --- stochastic setup time anticipation (z-Score) and explicit overtime (overtime penalty) --- consistently improves stochastic optimization across all utilization levels. Stochastic costs decrease significantly, stabilizing the planning--execution interaction, while the optimal flexibility level ($\tilde{T}$) shifts. At 80\% utilization, costs improve from 16,295 to 15,802, with optimal $\tilde{T}$ shifting from a rigid 12 to a more flexible 4 periods (z-Score 1.645, overtime 7.0). At 85\%, costs decrease from 18,819 to 18,139, with behavior reversing: the unadapted model relied on high flexibility ($\tilde{T}=1$), while the adapted model performs best with rigid planning ($\tilde{T}=12$, z-Score 0.842, overtime 7.0). The strongest effect occurs at 90\% utilization, where costs drop from 25,502 to 21,576---a 15.4\% improvement (z-Score 1.645, overtime 25.0)---with optimal $\tilde{T}$ shifting from a moderately flexible 7 to a rigid 12 periods.}

\begin{table}[htbp]
    \centering
    \resizebox{\textwidth}{!}{
        \setlength{\tabcolsep}{2pt}
        \begin{tabular}{c | ccccc | cccccc | cccccc | ccccccc | c}
            \hline
            \multicolumn{26}{c}{No safety stocks} \\
            \hline
            & \multicolumn{5}{c|}{MRP} & \multicolumn{6}{c|}{Deterministic} & \multicolumn{6}{c|}{Stochastic ($\tilde{T}=12$)} & \multicolumn{7}{c|}{Stochastic (Best $\tilde{T}$)} & \% $\Delta$ \\
            & Cost & LT & SS & SL & LSP & Cost & LT & SS & SL & Z & OT & Cost & LT & SS & SL & Z & OT & Cost & LT & SS & SL & $\tilde{T}$ & Z & OT & \\
            \hline
            80\% & 21,218 & 1 & 0 & 89.38\% & FOQ 1.5 & 37,110 & 1 & 0 & 43.84\% & 1.645 & 25.0 & 16,116 & 1 & 0 & 89.95\% & 0.842 & 7.0 & \textbf{15,802} & 1 & 0 & 89.86\% & 5 & 1.645 & 7.0 & -26\% \\
            85\% & 22,522 & 1 & 0 & 87.45\% & FOQ 1.5 & 38,100 & 1 & 0 & 44.87\% & 1.645 & 7.0 & \textbf{18,139} & 1 & 0 & 85.91\% & 0.842 & 7.0 & \textbf{18,139} & 1 & 0 & 85.91\% & 12 & 0.842 & 7.0 & -19\% \\
            90\% & 25,042 & 1 & 0 & 83.84\% & FOQ 1.5 & 39,573 & 1 & 0 & 46.17\% & 1.645 & 7.0 & \textbf{21,576} & 1 & 0 & 82.82\% & 1.645 & 25.0 & \textbf{21,576} & 1 & 0 & 82.82\% & 12 & 1.645 & 25.0 & -14\% \\
            \hline
            \multicolumn{26}{c}{With safety stocks} \\
            \hline
            & \multicolumn{5}{c|}{MRP} & \multicolumn{6}{c|}{Deterministic} & \multicolumn{6}{c|}{Stochastic ($\tilde{T}=12$)} & \multicolumn{7}{c|}{Stochastic (Best $\tilde{T}$)} & \% $\Delta$ \\
            & Cost & LT & SS & SL & LSP & Cost & LT & SS & SL & Z & OT & Cost & LT & SS & SL & Z & OT & Cost & LT & SS & SL & $\tilde{T}$ & Z & OT & \\
            \hline
            80\% & 20,969 & 1 & 0.2 & 92.81\% & FOQ 1.5 & \textbf{13,374} & 1 & 0.5 & 99.50\% & 1.645 & 7.0 & 15,059 & 1 & 0.1 & 92.26\% & 1.282 & 7.0 & 14,827 & 1 & 0.1 & 91.31\% & 3 & 1.645 & 7.0 & -64\% \\
            85\% & 22,522 & 1 & 0 & 87.45\% & FOQ 1.5 & \textbf{13,206} & 1 & 0.5 & 98.82\% & 0.0 & 25.0 & 16,587 & 1 & 0.2 & 88.85\% & 0.0 & 25.0 & 16,587 & 1 & 0.2 & 88.85\% & 12 & 0.0 & 25.0 & -41\% \\
            90\% & 25,042 & 1 & 0 & 83.84\% & FOQ 1.5 & \textbf{13,509} & 1 & 0.5 & 97.15\% & 1.645 & 7.0 & 19,771 & 1 & 0.6 & 88.18\% & 1.645 & 7.0 & 19,771 & 1 & 0.6 & 88.18\% & 12 & 1.645 & 7.0 & -46\% \\
            \hline
        \end{tabular}
    }
    \caption{Cost, parameter, service level (SL) overview for the production system with and without safety stocks. Comparison of MRP, deterministic optimization, and stochastic optimization ($\tilde{T}=12$ and Best $\tilde{T}$). The columns Z and OT represent the extended parameters z-Score and Overtime Value, respectively.}
    \label{tab:safety_stock_comparison_extended}
\end{table}

\paragraph{With‑safety‑stock setting}
\textcolor{black}{With safety stocks (Table \ref{tab:safety_stock_comparison_extended}), deterministic optimization dominates across all scenarios, as inventory buffers inherently mitigate much of the execution-level variability. Nonetheless, stochastic optimization benefits from execution-level adaptations in absolute terms: costs drop from 15,139 to 14,827 at 80\% (z-Score 1.645, overtime 7.0, $\tilde{T}=3$), from 17,371 to 16,587 at 85\% (z-Score 0.0, overtime 25.0, $\tilde{T}=12$), and from 21,825 to 19,771 at 90\% (z-Score 1.645, overtime 7.0, $\tilde{T}=12$). This confirms that setup-time feedback and limited overtime improve stochastic performance, enabling more rigid planning horizons ($\tilde{T}=12$). By contrast, deterministic optimization benefits only marginally, with costs improving slightly to 13,374 (80\%), 13,206 (85\%), and 13,509 (90\%). Relying heavily on safety stock (SS=0.5) to absorb variability, it uses overtime merely as a reactive feasibility buffer without altering the production plan. Lacking variability anticipation, deterministic optimization cannot strategically exploit the z-Score and overtime, making these adaptations primarily advantageous for the stochastic model.}

\textcolor{black}{
Turning to specific metrics: cumulative demand and production trajectories with and without safety stock appear in Figures \ref{fig:cumulative_flow_diagramm_no_safety_stock} and \ref{fig:cumulative_flow_diagramm_with_safety_stock} of the Appendix for one selected item, while Table \ref{tab:mixed_system_complete_safety_stocks} reports comprehensive performance metrics. Table \ref{tab:mixed_system_complete_safety_stocks} confirms deterministic optimization achieves the lowest costs at all utilization levels (13,374/13,206/13,509 at 80\%/85\%/90\%) for the 95\% target service level base case and maintains superiority across additional sensitivity scenarios (90\% and 98\% service levels, FGI cost factor of 4). The cost factor sensitivity analysis shows that varying these ex-post parameters acts as linear multipliers on the underlying operational logic, leaving relative rankings and conclusions unchanged. This dominance is driven by near-zero End Item lateness ($\mathit{Lateness}_{EI} \approx 0.0358$) and very short market lead times ($\mathit{Market~LT} \approx 0.0120$), indicating inventory buffering effectively absorbs execution-level variability. Although deterministic plans with safety stock result in higher component inventory levels (e.g., 922.43 at the 90\% safety stock level) compared with plans without safety stock, they reduce lateness and tardiness costs to only 2\%--6\% of total costs, which more than compensates for the additional inventory.}

\begin{table}[h]
\resizebox{\textwidth}{!}{
    \centering
    \setlength{\tabcolsep}{3pt}
    \begin{tabular}{l|ccc|cccc|cc}
    \hline
    \multicolumn{10}{c}{Mixed Production System - Comprehensive Performance Overview} \\
    \hline 
    & \multicolumn{3}{c|}{\textbf{Operative Performance (Base Case 95\%)}} & \multicolumn{4}{c|}{\textbf{Ex-post Overall Costs (CU) Sensitivity}} & \multicolumn{2}{c}{\textbf{Avg. Inventory}} \\
    \textbf{Method} & \textbf{SL (\%)} & Lateness EI & Market LT & \textbf{Base ($c_b$=38)} & $c_b$=19 & $c_b$=98 & FGI=4 & End Item & Comp. \\
    \hline
    \multicolumn{10}{l}{\textbf{Utilization 80\%}} \\
    \hline
    MRP           & 92.81\% & 0,0358 & 0,0120 & 20.969 & 20.203 & 23.390 & 27.327 & 856 & 1692 \\
    Deterministic & 99.50\% & 0,0000 & 0,0026 & \textbf{13.374} & \textbf{13.268} & \textbf{13.707} & \textbf{17.996} & 593 & 925  \\
    Stoch. T=12   & 92.26\% & 0,0262 & 0,0218 & 15.059 & 14.147 & 17.937 & 18.797 & 512 & 1198 \\
    Stoch. Best T & 91.31\% & 0,0226 & 0,0206 & 14.827 & 14.026 & 17.358 & 18.603 & 521 & 1196 \\
    \hline
    \multicolumn{10}{l}{\textbf{Utilization 85\%}} \\
    \hline
    MRP           & 87.45\% & 0,0783 & 0,0543 & 22.522 & 20.355 & 29.365 & 27.755 & 773 & 1668 \\
    Deterministic & 98.82\% & 0,0358 & 0,0120 & \textbf{13.206} & \textbf{13.074} & \textbf{13.623} & \textbf{17.685} & 856 & 1692 \\
    Stoch. T=12   & 88.85\% & 0,0569 & 0,0396 & 16.587 & 15.055 & 21.427 & 20.579 & 574 & 1210 \\
    Stoch. Best T & 88.85\% & 0,0569 & 0,0396 & 16.587 & 15.055 & 21.427 & 20.579 & 574 & 1210 \\
    \hline
    \multicolumn{10}{l}{\textbf{Utilization 90\%}} \\
    \hline
    MRP           & 83.84\% & 0,1454 & 0,1052 & 25.042 & 21.285 & 36.904 & 29.942 & 772 & 1566 \\
    Deterministic & 97.15\% & 0,0018 & 0,0095 & \textbf{13.509} & \textbf{13.117} & \textbf{14.747} & \textbf{17.849} & 597 & 922  \\
    Stoch. T=12   & 88.18\% & 0,1017 & 0,0464 & 19.771 & 17.864 & 25.793 & 25.864 & 855 & 1234 \\
    Stoch. Best T & 88.18\% & 0,1017 & 0,0464 & 19.771 & 17.864 & 25.793 & 25.864 & 855 & 1234 \\
    \hline
    \end{tabular}
}
\caption{Performance comparison including operative delivery metrics (Service Level and Lateness) and specific sensitivity analysis of the ex-post overall costs for different production control methods in the setting with safety stocks.}
\label{tab:mixed_system_complete_safety_stocks}
\end{table}
\textcolor{black}{Stochastic optimization also benefits from execution-level adaptations in absolute terms but remains more expensive than deterministic optimization in the safety-stock setting. As shown in Table \ref{tab:mixed_system_complete_safety_stocks}, the best stochastic solutions reduce overall costs to 14,827 at 80\% utilization, 16,587 at 85\%, and 19,771 at 90\%. However, stochastic optimization still exhibits higher $\mathit{Lateness}_{EI}$ (up to 0.1017 at 90\%) and longer market lead times than deterministic planning. This reflects the fact that the anticipative features of stochastic optimization provide limited additional value once sufficient deterministic inventory buffers are already present. In summary, the Tables \ref{tab:safety_stock_comparison_extended} and \ref{tab:mixed_system_complete_safety_stocks} confirm that safety stock fundamentally changes system behavior: inventory buffering stabilizes execution so effectively that deterministic optimization benefits disproportionately, while stochastic anticipation plays a secondary role. By contrast, the main gains of stochastic optimization and execution-level adaptations (z-Score and overtime) arise in the absence of safety stock, where these mechanisms become critical for handling tight capacities and shop-floor noise.}

\subsubsection{Summary use case evaluation}
\textcolor{black}{To answer RQ4 and RQ5, the evaluation of the complex use cases  demonstrates the substantial advantages of optimization-based planning over standard MRP, yielding cost reductions of up to 68\%. The optimal planning model depends heavily on inventory policy and system configuration: without safety stocks, stochastic optimization with execution-level adaptations (overtime and setup time anticipation) provides the most robust schedules; with safety stocks, deterministic optimization benefits disproportionately, leveraging inventory buffers to absorb shop-floor noise. Production flexibility ($\tilde{T}$) must be tailored to system structure: simple convergent systems thrive on high flexibility ($\tilde{T}=1$), while complex divergent and mixed systems require rigid planning horizons at high utilization to maintain coordination stability.}

\textcolor{black}{These findings align with prior literature: the dominance of buffered deterministic planning is consistent with \citet{Bitran.1984}, \citet{Inderfurth.2009}, and \citet{Mula.2006}, who show that safety stocks effectively substitute for stochastic anticipation, while the high value of stochastic optimization in buffer-free settings reflects the theoretical results on the value of the stochastic solution \citep{Birge.1982, Huang.2009}. However, prior work has studied these effects largely in isolation --- safety stock policies within deterministic MRP or stochastic programming in stylized models. Our contribution bridges this gap by embedding both paradigms within a coupled planning--execution simulation under realistic shop-floor conditions and across diverse production structures, providing systematic evidence on how execution-level adaptations, inventory buffering, and production structure jointly shape relative planning performance --- an interaction not comprehensively investigated before. These insights provide the foundation for the final conclusions of this study.}

\section{Conclusion}
\label{sec:conclusion}

In this work, we combine a \Rev{two-stage} stochastic optimization model for multi-item multi-echelon capacitated lot sizing with discrete-event simulation in order to generate a rolling horizon production planning framework. Making use of stochastic optimization allows to explicitly consider customer demand fluctuations in the lot sizing process. \Rev{Within the stochastic model we introduce a parameter that allows to consider the possibility to flexibly adjust production quantities in later periods and hence more accurately represents the underlying decision process.} We compare the stochastic planning approach to a deterministic optimization model using expected demands, as well as to a standard MRP approach. We evaluate the different planning methods by means of the developed discrete-event simulation framework and report the resulting overall costs. 

\textcolor{black}{Regarding the research question RQ1 of how the performance of stochastic optimization compares to deterministic optimization and standard MRP in rolling horizon seeting, our results show that under tight resource restrictions MRP cannot anticipate capacity bottlenecks caused by irregular demand patterns, even when short-term demands remain unchanged. In these settings, deterministic optimization already improves plan stability, while stochastic optimization performs best by smoothing capacity utilization.}

\textcolor{black}{Analyzing customer update behaviors in RQ2 reveals that, results show that higher update frequencies (as in Customer Types B and C) increasingly challenge MRP and deterministic optimization, while stochastic optimization remains robust by explicitly anticipating short-term demand changes.}

\textcolor{black}{With respect to the research question RQ3 on how the number of scenarios affects cost-effective stochastic production plans, our results show that solution quality improves considerably when increasing the scenario set up to around 30, while further increases offer only minor additional benefits relative to the computational effort.}

\textcolor{black}{For research question RQ4 targeting varying shop load, the results show that higher shop-floor loads and increasing production system complexity strengthen the performance advantage of stochastic optimization, while deterministic models lose effectiveness as capacity becomes tight.}

\textcolor{black}{Finally, regarding RQ5, integrating probabilistic setup-time adaptations 
and soft overtime constraints substantially stabilizes the planning--execution 
horizon. In highly congested, unbuffered settings, this mechanism recovers 
schedule feasibility, reducing stochastic costs by up to 15.4\% at 90\% 
utilization. However, this anticipative value diminishes once explicit safety 
stocks are introduced, as physical inventory buffers inherently absorb 
execution friction.}

\textcolor{black}{For future work, we plan to extend our framework to a multi-stage stochastic model that explicitly accounts for rolling-horizon adjustments in later periods. This future research will employ sophisticated parameter tuning to analyze safety buffer behavior in stochastic optimization. We also intend to investigate hybrid structures where end items utilize optimization models while components follow standard MRP logic, examining the impact of varying Time-Between-Orders (TBO) on capacity smoothing. Additionally, incorporating sequence-dependent setups and setup carry-overs represents a promising avenue, especially for systems with high-frequency micro-periods or multi-period production runs.}

\section*{Funding}
This research was funded in whole, or in part, by the Austrian Science Fund (FWF) [10.55776/P32954]. For the purpose of open access, the author has applied a CC BY public copyright licence to any Author Accepted Manuscript version arising from this submission.

\section*{Data availability statement}
The data that support the findings of this study are available in Zenodo at 10.5281/zenodo.10686729    upon reasonable request to Wolfgang Seiringer.

\bibliography{MRP-SimOpt}

\newpage
\appendix

\section{Planning and interaction requirements}
\label{app:planning_interaction}
Rolling horizon planning is dynamic, involving several critical steps \citep{figueira_hybrid_2014, Sahin.2013}. The process begins by starting the simulation with a limited runtime of $n$ periods. During the simulation, a rolling horizon planning window, which captures a timespan of $T$ periods  from the current simulation time $\bar{t}$, is iteratively shifted by one discrete time period. In this context, $T$ denotes the length of the planning window. A shift of the rolling horizon window is performed by moving the observed interval from $(\bar{t},\bar{t}+T)$ to $(\bar{t}+1,\bar{t}+1+T)$. Each shift or increase of $\bar{t}$ to $\bar{t}+1$ triggers a new planning cycle, until the current simulation time $\bar{t}$ is equal to the runtime limit $n$. 

After a new planning cycle has started, new customer orders are generated and existing orders are updated according to the forecast evolution model outlined in Section \ref{forecastEvolutionModel}. The demand forecast of an order for item $i$ and due date $t$ is denoted by $D_{itb}$, assuming that the current simulation time is $b$ periods before the due date, i.e. $t-\bar{t}=b$. 
After generating and updating customer demands, the simulation provides two types of parameters to the optimization component. First, general information on the production system setting is required. This comprises information on the components $\mathcal{I}_c$ and end items $\mathcal{I}_e$, as well as resources $\mathcal{K}$ and respective resource requirements of items ($\mathcal{I}_k$ and $\mathcal{K}_i$). Furthermore, BOM structure ${R}_{ij}$ and time related parameters, e.g. lead times ${L}_i$, setup times ${t}_{ik}$ and processing times ${p}_{ik}$, need to be provided. Finally, also cost-related parameters, e.g. setup ${s}_{ik}$, processing ${v}_{i}$, holding ${h}_{i}$, backlog ${b}_i$ and lost sales costs ${e}_i$, are part of the required general information and need to be passed to the optimization component. These parameters stay constant during the whole simulation runtime. 

Next to the general information, the optimization component also requires dynamic parameters that change during simulation. These include the inventory at hand $\hat{I}_{i0}$ and information on arriving orders during the planning window $\hat{I}_{it}$, as well as available resource capacity ${C}_{kt}$ and finally, the updated customer demand forecasts ${D}_{itb}$. To provide the initial inventory $\hat{I}_{i0}$, the current on stock quantity of item $i$ at the current simulation time is reported. To identify the production orders $\hat{I}_{it}$ of item $i$ that will arrive in period $t$, all waiting and processing orders with planned end $t$ are selected and the respective order quantities are summed. To compute the available capacity ${C}_{kt}$ of a resource $k$ in period $t$,  all planned and running production orders in the simulation are selected and their required capacity is subtracted from the initially available capacity of the upcoming period. For currently running production orders (at time $\bar{t}$), only the remaining capacity requirement between the current simulation time and the planned end date is taken into consideration. For waiting production orders, the full requirement is subtracted. In case the summed capacity requirement of the production orders in the simulation exceeds the capacity of a resource in the upcoming period, capacity of subsequent periods is consumed, meaning that $C_{kt}$ is reduced,  respectively. Finally, the latest demand information for customer orders within the planning horizon is provided to the optimization component. Depending on the optimization approach, different parameters are necessary. For the deterministic optimization model it is sufficient to report the most recent demand forecasts $D_{itb}$, which are used as the deterministic demands $D_{it}$ in the model. The scenario-based stochastic optimization on the other hand, makes use of different scenarios $\omega \in \Omega$ and the respective scenario demands $D_{it}^{\omega}$ to generate production plans that perform well in the face of uncertain customer demand. In order to generate meaningful demand scenarios, we use 
historical demand forecasts $D_{it'b}$, where the due date $t'$ was already prior to the current simulation time $\bar{t}$, i.e. $t'<\bar{t}$,
to estimate how much the actually realized demand at the due date will differ from its prediction. We compare the predicted quantity to the actual quantity when the order is due and track this deviation during the whole demand information horizon. We collect these deviations during the warm-up phase $z$ of the simulation, i.e. $0 \leq \bar{t} \leq z$, and calculate the mean deviation for each combination of end item $i$ and distance to due date $b$. This allows us to estimate the standard deviation $\sigma_{ib}$ of a truncated normal distribution, which is later used to draw the scenario demands $D_{it}^{\omega}$ for the stochastic optimization model. Assuming that the due date $t$ is reached in $b$ periods, the respective mean of the truncated normal distribution is modeled as the most current prediction $D_{itb}$, which represents the most up-to-date information available. To summarize, the scenario demands $D_{it}^{\omega}$ for end item $i$ at due date $t$ used in the stochastic optimization model are drawn from a truncated normal distribution with mean $D_{itb}$ and standard deviation $\sigma_{ib}$, assuming that period $t$ is $b$ periods from the current simulation time. For a more detailed explanation of this procedure we refer to \cite{Altendorfer.Felberbauer.2023}. After all necessary general and dynamic parameters are provided to the optimization component, the chosen model can be formulated and solved. The simulation environment is required to await the resolution of the optimization problem. After solving the model, the optimization component returns the optimal decision variables, i.e. the production quantities ${Q}_{it}$ to the simulation. Production quantities having a period index corresponding to the current simulation time $\bar{t}$ are used to generate production orders in the simulation, e.g. for $Q_{it} = 200$, a production order of 200 pieces for item $i$ is generated if $t=\bar{t}$. The end date of each created production order of item $i$ is set to the sum of the planned start date and the respective planned lead time $L_i$. Finally, the orders are released and processed on the shop-floor, where the released production orders of previous periods are processed as well. The current period is simulated until the beginning of the next discrete period, which triggers the next planning cycle. 

\section{MRP baseline}
\label{app:mrp_basline}
\begin{figure}[ht]
    \centering
    \includegraphics[width=\textwidth]{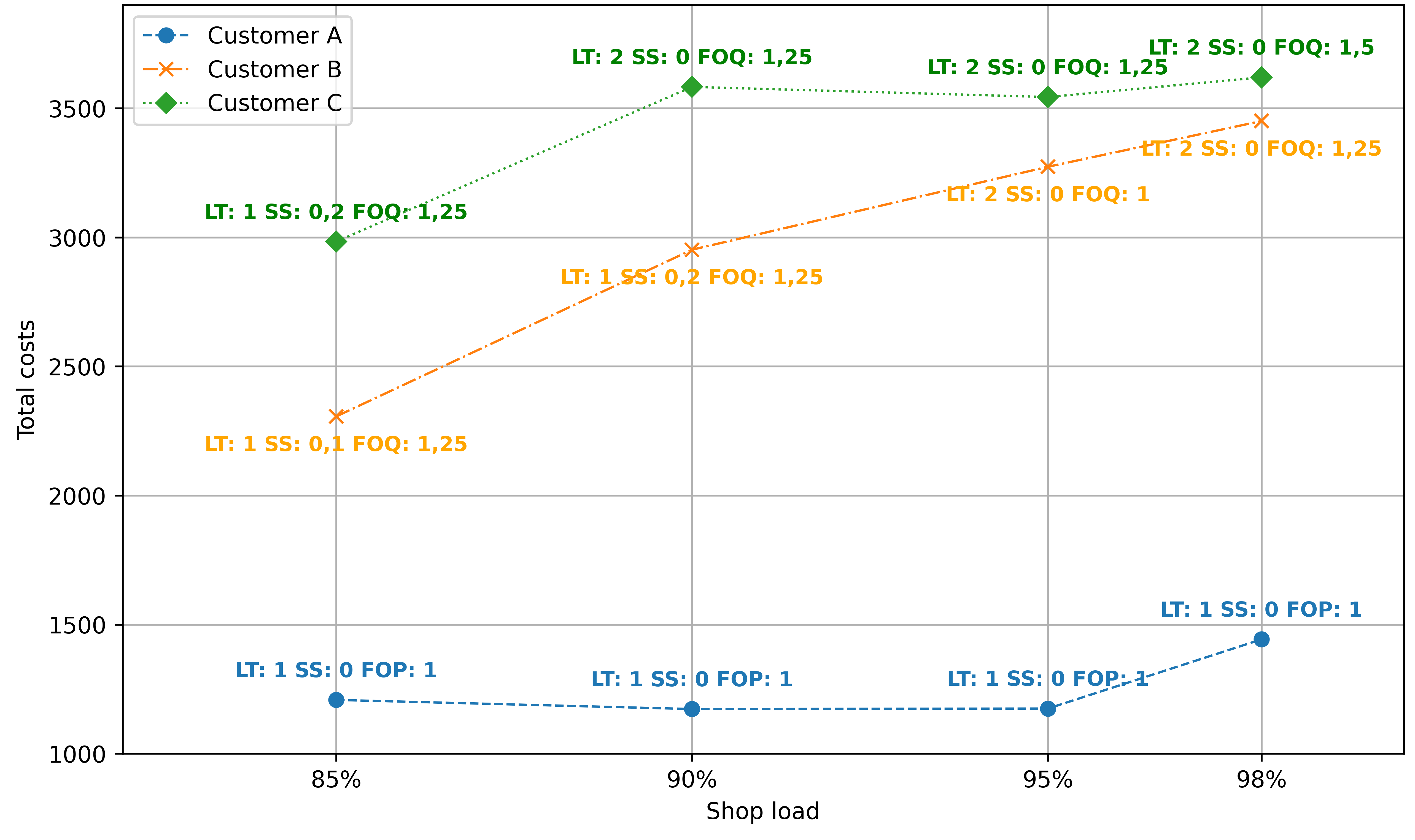}
    \caption{MRP baseline: cost comparison for MRP with medium demand variation of 7.5\%}
    \label{fig:mrp_baseline_with_annotations}
\end{figure}

While the MRP data presented in Figure \ref{fig:mrp_baseline_with_annotations} does not introduce novel MRP-specific discoveries, they do offer valuable perspectives on planning outcomes and validate the accuracy of the MRP simulation. \textcolor{black}{To ensure a highly competitive baseline, the standard MRP approach was thoroughly tuned based on extensive preliminary studies, evaluating various job sorting sequences, planning horizons, and over 50,000 parameter combinations to identify the best-performing configurations.} Within Figure \ref{fig:mrp_baseline_with_annotations}, the minimum MRP overall costs associated with the simulated shop loads are illustrated. Each curve symbolizes distinct customer behaviors (reliable Customer Type A, volatile Customer Type B, nervous Customer Type C) subjected to a medium demand variation of 7.5\%. Based on these behaviors, Customer Type A  invariably results in the lowest overall costs across all shop loads. 

The optimal planning parameters for Customer Type A, which result in the minimum total costs, have been identified as LT (Lead Time) \(= 1\), SS (Safety Stock) \(= 0\), and FOP (Fixed Order Period) \(= 1\). This strategy is effective due to Customer Type A's deterministic demand behavior, which enables the production system to counteract stochastic effects without needing safety stock. When the shop load reaches \(98\%\), there is a slight increase in overall costs, as the higher production lead time causes increased tardiness. 

For Customer Type B, employing FOQ consistently results in the minimum overall costs, effectively mitigating the impact of a higher shop load. Specifically, at shop loads of \(85\%\) and \(90\%\), the most cost-effective approach involves a modest safety stock combined with LT \(= 1\) and FOQ \(= 1.25\%\). At higher shop loads of \(95\%\) and \(98\%\), the best planning performance is achieved by increasing the lead time to 2 periods instead of using safety stock.

A similar trend is observed for the nervous Customer Type C. For a shop load of \(85\%\), combining a lead time of 1 with safety stock attains the lowest overall costs. For higher shop loads, however, it is more effective to replace safety stock with an increased lead time. 

In conclusion, these identified parameter combinations are meaningful from a production logistics perspective. For reliable Customer Type A  minor variations from the long-term forecast can still be efficiently managed within a lead time of 1, without safety stock, and by adopting a lot-for-lot approach. Similarly, for Customer Types B and C, who exhibit more volatile and fluctuating demand patterns, meaningful planning parameters are employed. The higher shop load necessitates compensation through either a safety stock or a longer lead time. An increased lead time helps avoid production delays caused by larger lot sizes, while a small safety stock can effectively buffer against demand fluctuations. \textcolor{black}{This aligns with MRP literature: safety lead times mitigate timing/capacity uncertainties, while safety stocks buffer demand variations \citep{Whybark.1976, v.d.Guide.2000, Dolgui.2008}.}

\section{Additional figures}

\begin{figure}[!ht]
    \centering
        \includegraphics[width=\linewidth]{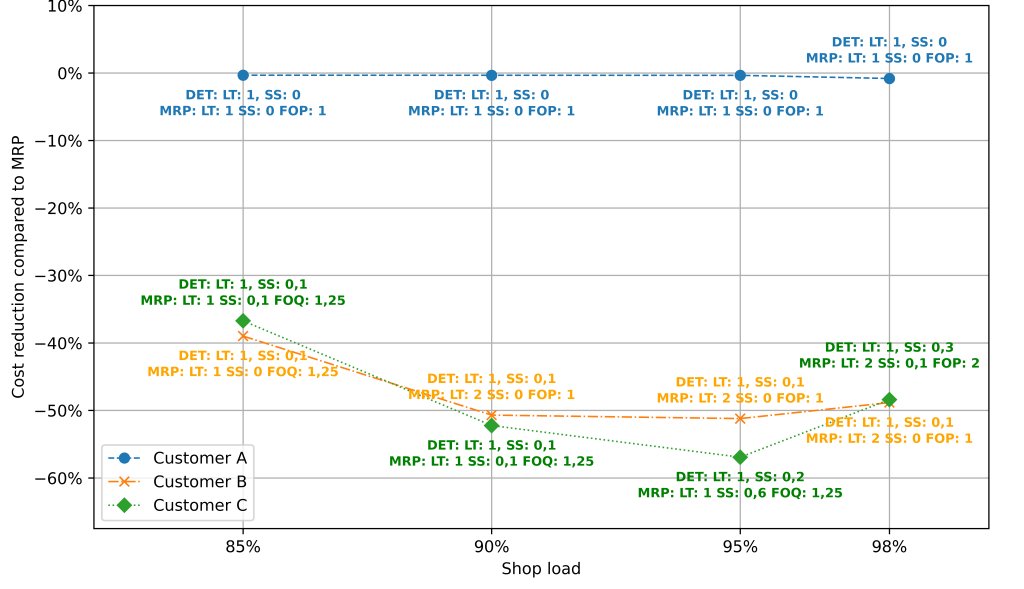}
        \caption{Cost comparison of deterministic optimization to MRP in a setting with low demand variation (2.5\%) for different shop loads}
        \label{fig:best_performing_deterministic_low_variation}
\end{figure}

\begin{figure}[!ht]
    \centering
        \includegraphics[width=\linewidth]{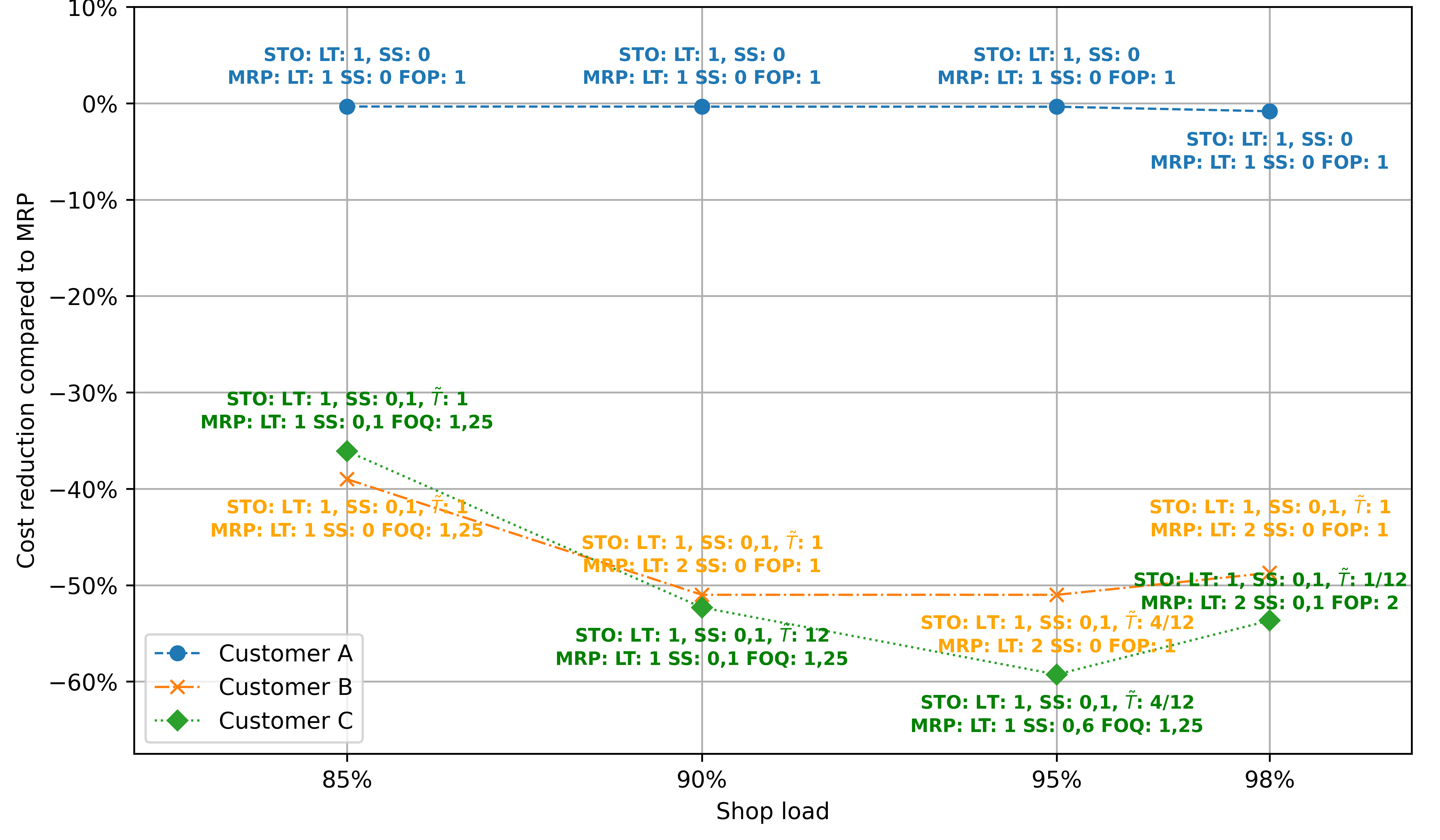} 
        \caption{Cost comparison of stochastic optimization (30 scenarios) to MRP in a setting with low demand variation (2.5\%) for different shop loads}
        \label{fig:best_performing_stochastic_low_variation}
\end{figure}

\begin{figure}[!ht]
    \centering
        \includegraphics[width=\linewidth]{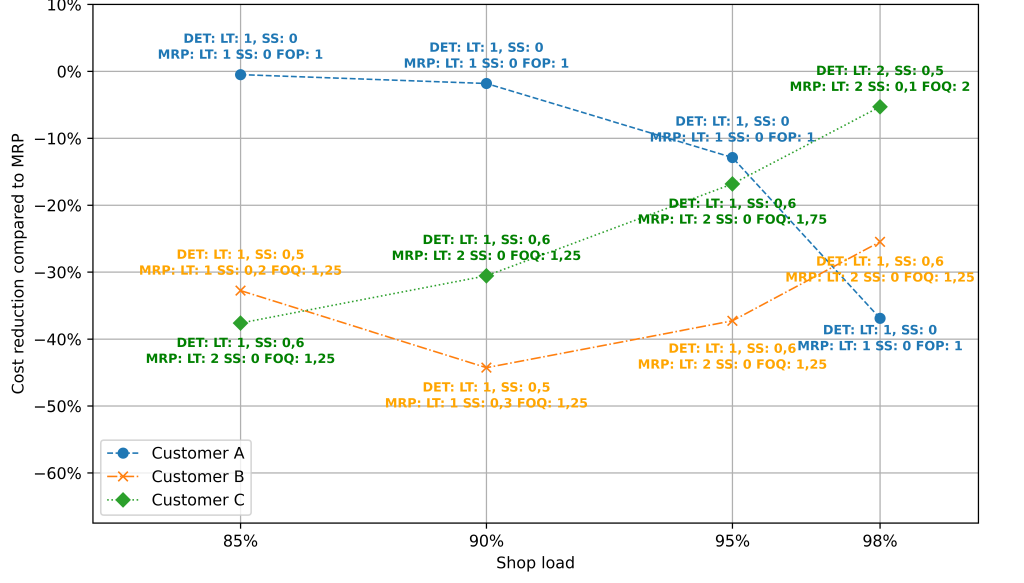}
        \caption{Cost comparison of deterministic optimization to MRP in a setting with high demand variation (12.5\%) for different shop loads}
        \label{fig:best_performing_deterministic_high_variation}
\end{figure}

\begin{figure}[!ht]
    \centering
        \includegraphics[width=\linewidth]{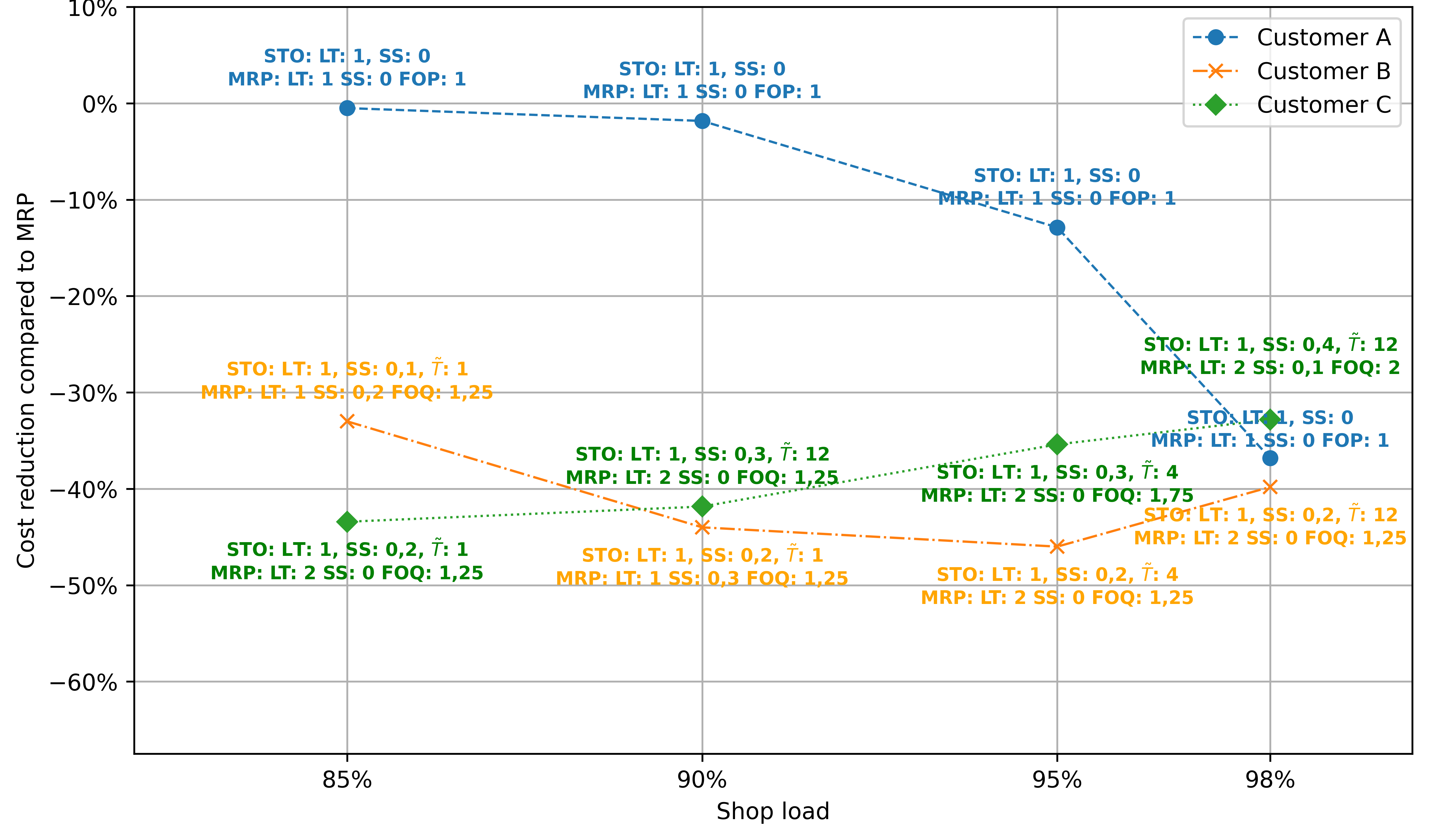} 
        \caption{Cost comparison of stochastic optimization (30 scenarios) to MRP in a setting with high demand variation (12.5\%) for different shop loads}
        \label{fig:best_performing_stochastic_high_variation}
\end{figure}

\clearpage

\begin{figure}[!ht]
    \centering
    \includegraphics[width=\textwidth]{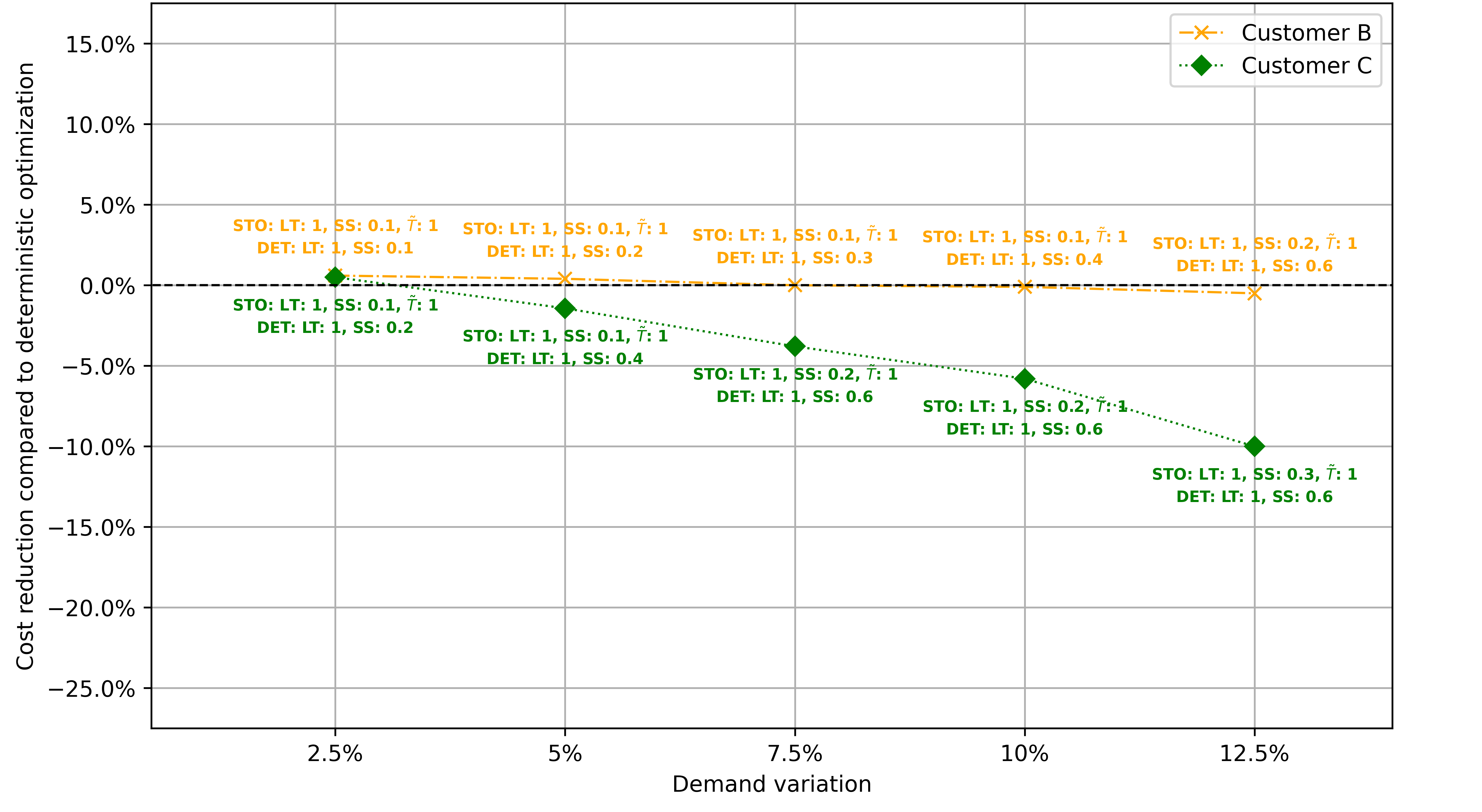}
    \caption{Cost comparison of stochastic optimization (30 scenarios) to deterministic optimization with a utilization of 85\% for different demand variations}
    \label{fig:forecast_uncertainty_85}
\end{figure}

\begin{figure}[ht]
    \centering
    \includegraphics[width=\textwidth]{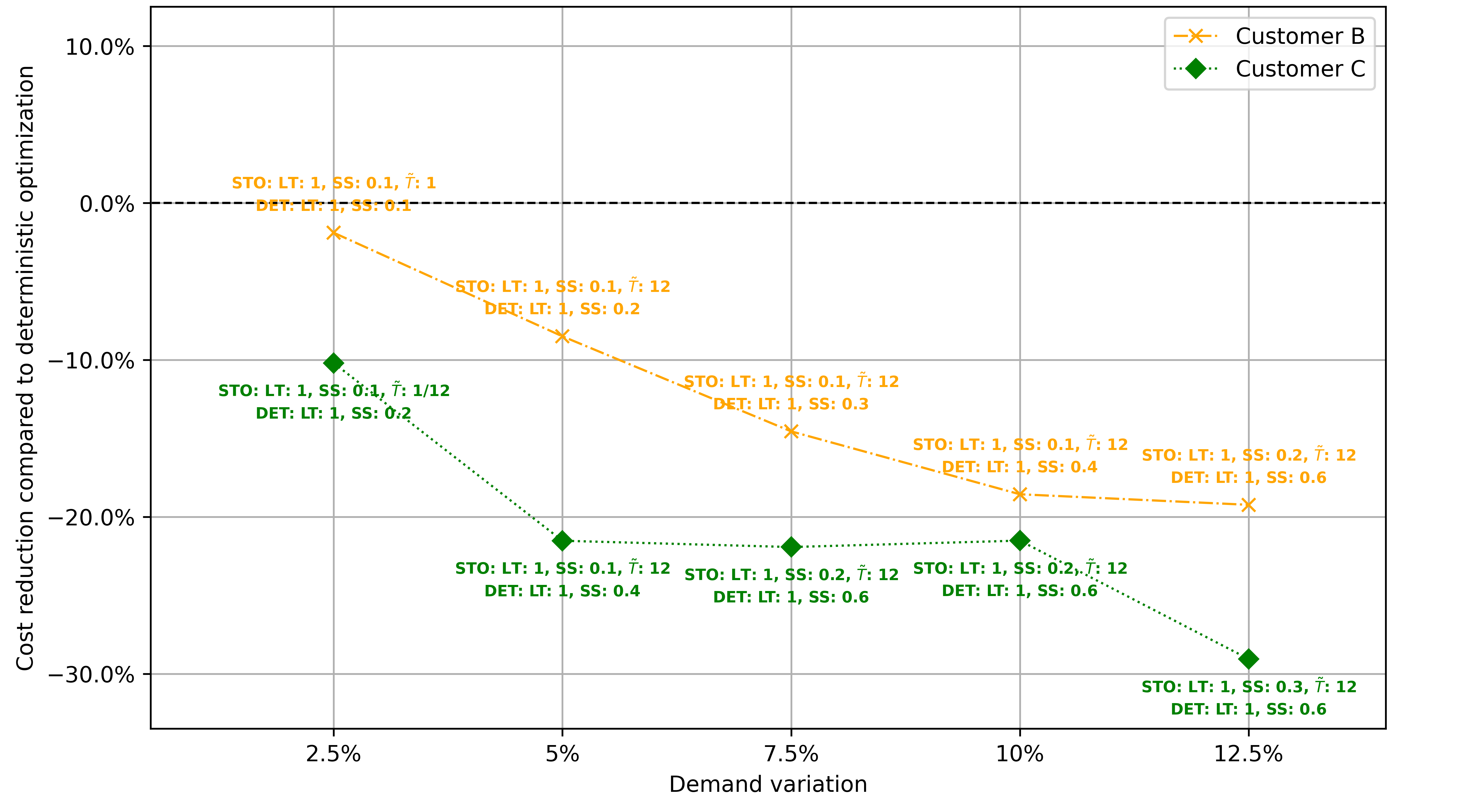}
    \caption{Cost comparison of stochastic optimization (30 scenarios) to deterministic optimization with a utilization of 98\% for different demand variations}
    \label{fig:forecast_uncertainty_98}
\end{figure}

The low utilization case Figure \ref{fig:forecast_uncertainty_85} demonstrates that deterministic optimization with safety stocks is sufficient for Customer Type B if enough resource capacity is available. It outperforms the stochastic approach for all demand variations. For Customer Type C solving the stochastic model is advantageous and cost improvements of up to 10\% over the deterministic version are possible. In high utilization environments of 98\% (Figure \ref{fig:forecast_uncertainty_98}), stochastic optimization outperforms the deterministic model for both Customer Types B and C and over all demand variations. While improvements of up to 20\% are possible for Customer Type B, a cost reduction of up to 30\% can be reached for Customer Type C.

\clearpage
\section{General overview}
\label{sec:General_Overview_Appendix}

Table \ref{tab:general_overview_combined_table} presents the performance, optimal parameters, and relative cost differences ($\%\Delta$) compared to standard MRP across all evaluated settings, highlighting the best models in bold. As expected, average overall costs increase with higher utilization due to capacity bottlenecks.

\begin{itemize}
    \item \textbf{Customer Type A (Stable):} At moderate utilizations (85\% and 90\%), optimization yields no advantage over MRP. Because demand remains stable, the stochastic model is equivalent to the deterministic version (hence only deterministic results are reported). However, at high utilizations (95\% and 98\%) with high demand variation, optimization mitigates short-term bottlenecks, reducing costs by up to 13\% (for 95\% utilization) and 37\% (for 98\% utilization).
    
    \item \textbf{Customer Type B (Short-term updates):} At lower utilizations, deterministic models using safety stocks are most efficient, achieving cost reductions of up to 50\% over MRP. At higher utilizations, tight capacities restrict the creation of large safety stocks, causing a decline in deterministic performance. Here, stochastic optimization becomes the superior choice, especially for large demand variations.
    
    \item \textbf{Customer Type C (Frequent updates):} Representing the highest uncertainty, stochastic optimization is superior across all utilizations and demand variations. By explicitly anticipating future demand realizations, it reduces costs by over 50\% compared to MRP, whereas deterministic models fail despite large safety stocks.
\end{itemize}

Moreover, regarding the stochastic model, assuming flexible production quantities is advantageous at lower resource utilizations. Conversely, at higher utilizations, a conservative approach of assuming them to be fixed is beneficial. Assuming full flexibility can be overly optimistic in a rolling horizon setting, as it underestimates tight resource constraints when decisions are actually made period by period.

\begin{sidewaystable}
\resizebox{\textwidth}{!}{
    \centering
    \begin{tabular}{|ll|llllllll|r|llllllll|r|}
    \hline
        ~ & ~ & \multicolumn{9}{c|}{Utilization 85\%} & \multicolumn{9}{c|}{Utilization 90\%} \\ \hline
        Type & Variation & \multicolumn{2}{c}{MRP} & \multicolumn{2}{c}{Deterministic} & \multicolumn{2}{c}{Stochastic} & \multicolumn{2}{c|}{Stochastic} & \%$\Delta$ & \multicolumn{2}{c}{MRP} & \multicolumn{2}{c}{Deterministic} & \multicolumn{2}{c}{Stochastic} & \multicolumn{2}{c|}{Stochastic} & \%$\Delta$\\
        ~ & ~ & \multicolumn{2}{c}{(LT,SS,LSP)} & \multicolumn{2}{c}{(LT,SS)} & \multicolumn{2}{c}{($\tilde{T}$:12) (LT,SS)}& \multicolumn{2}{c|}{($\tilde{T}$:1) (LT,SS)}& &  \multicolumn{2}{c}{(LT,SS,LSP)} & \multicolumn{2}{c}{(LT,SS)} & \multicolumn{2}{c}{($\tilde{T}$:12) (LT,SS)} & \multicolumn{2}{c|}{($\tilde{T}$:1) (LT,SS)} &\\ \hline
        A\textsuperscript{1} & 2.5\% & 1209 & (1,0,P: 1) & \textbf{1205} & (1,0) & \multicolumn{2}{c}{-} & \multicolumn{2}{c|}{-} & 0\% & 1174 & (1,0,P: 1) & \textbf{1170} & (1,0) & \multicolumn{2}{c}{-} & \multicolumn{2}{c|}{-} & 0\% \\ 
        A\textsuperscript{1} & 5.0\% & 1209 & (1,0,P: 1) & \textbf{1205} & (1,0) & \multicolumn{2}{c}{-} & \multicolumn{2}{c|}{-} & 0\% & 1173 & (1,0,P: 1) & \textbf{1169} & (1,0) & \multicolumn{2}{c}{-} & \multicolumn{2}{c|}{-} & 0\% \\ 
        A\textsuperscript{1} & 7.5\% & 1208 & (1,0,P: 1) & \textbf{1204} & (1,0) & \multicolumn{2}{c}{-} & \multicolumn{2}{c|}{-} & 0\% & 1173 & (1,0,P: 1) & \textbf{1169} & (1,0) & \multicolumn{2}{c}{-} & \multicolumn{2}{c|}{-} & 0\% \\ 
        A\textsuperscript{1} & 10.0\% & 1207 & (1,0,P: 1) & \textbf{1203} & (1,0) & \multicolumn{2}{c}{-} & \multicolumn{2}{c|}{-} & 0\% & 1178 & (1,0,P: 1) & \textbf{1169} & (1,0) & \multicolumn{2}{c}{-} & \multicolumn{2}{c|}{-} & -1\% \\ 
        A\textsuperscript{1} & 12.5\% & 1208 & (1,0,P: 1) & \textbf{1202} & (1,0) & \multicolumn{2}{c}{-} & \multicolumn{2}{c|}{-} & 0\% & 1195 & (1,0,P: 1) & \textbf{1173} & (1,0) & \multicolumn{2}{c}{-} & \multicolumn{2}{c|}{-} & -2\% \\ \hline
        B\textsuperscript{2} & 2.5\% & 2179 & (1,0,Q: 1.25) & \textbf{1330} & (1,0.1) & 1410 & (1,0.1) & 1338 & (1,0) & -39\% & 2626 & (2,0,P: 1) & \textbf{1295} & (1,0.1) & 1375 & (1,0.1) & 1306 & (1,0) & -51\% \\ 
        B\textsuperscript{2} & 5.0\% & 2212 & (1,0.1,Q: 1.25) & \textbf{1450} & (1,0.2) & 1505 & (1,0.1) & 1456 & (1,0) & -34\% & 2901 & (1,0.1,Q: 1.25) & \textbf{1419} & (1,0.2) & 1469 & (1,0.1) & 1427 & (1,0) & -51\% \\ 
        B\textsuperscript{2} & 7.5\% & 2306 & (1,0.1,Q: 1.25) & \textbf{1564} & (1,0.3) & 1628 & (1,0.1) & \textbf{1564} & (1,0) &-32\% & 2952 & (1,0.2,Q: 1.25) & \textbf{1542} & (1,0.3) & 1593 & (1,0.1) & 1549 & (1,0) & -48\% \\ 
        B\textsuperscript{2} & 10.0\% & 2455 & (1,0.2,Q: 1.25) & 1680 & (1,0.4) & 1748 & (1,0.1) & \textbf{1677} & (1,0) & -32\% & 3063 & (1,0.2,Q: 1.25) & \textbf{1675} & (1,0.4) & 1720 & (1,0.1) & 1681 & (1,0) & -45\% \\ 
        B\textsuperscript{2} & 12.5\% & 2675 & (1,0.2,Q: 1.25) & 1798 & (1,0.5) & 1880 & (1,0.1) & \textbf{1788} & (1,0) & -33\% & 3263 & (1,0.3,Q: 1.25) & 1818 & (1,0.5) & 1857 & (1,0.2) & \textbf{1801} & (1,0.1) & -44\% \\ \hline
        C\textsuperscript{3} & 2.5\% & 2202 & (1,0.1,Q: 1.25) & \textbf{1394} & (1,0.1) & 1408 & (1,0.1) & 1401 & (1,0) & -37\% & 2877 & (1,0.1,Q: 1.25) & 1375 & (1,0.1) & \textbf{1372} & (1,0.1) & 1390 & (1,0) & -52\% \\ 
        C\textsuperscript{3} & 5.0\% & 2423 & (1,0.1,Q: 1.25) & 1560 & (1,0.3) & 1538 & (1,0.1) & \textbf{1536} & (1,0.1) & -37\% & 3174 & (1,0.2,Q: 1.25) & 1603 & (1,0.3) & \textbf{1525} & (1,0.1) & \textbf{1525} & (1,0.1) & -52\% \\ 
        C\textsuperscript{3} & 7.5\% & 2983 & (1,0.2,Q: 1.25) & 1780 & (1,0.4) & 1712 & (1,0.2) & \textbf{1675} & (1,0.1) & -43\% & 3583 & (2,0,Q: 1.25) & 1964 & (1,0.5) & 1765 & (1,0.2) & \textbf{1764} & (1,0.1) & -51\% \\ 
        C\textsuperscript{3} & 10.0\% & 3613 & (1,0.3,Q: 1.25) & 1970 & (1,0.5) & 1875 & (1,0.2) & \textbf{1864} & (1,0.1) & -48\% & 3628 & (2,0,Q: 1.25) & 2284 & (1,0.5) & \textbf{1973} & (1,0.2) & 2024 & (1,0.1) & -46\% \\ 
        C\textsuperscript{3} & 12.5\% & 3619 & (2,0,Q: 1.25) & 2258 & (1,0.6) & 2034 & (1,0.2) & \textbf{2026} & (1,0.2) & -44\% & 3738 & (2,0,Q: 1.25) & 2596 & (1,0.6) & \textbf{2181} & (1,0.3) & 2259 & (1,0.2) & -42\% \\ \hline
        Average & ~ & 2181 & ~ & 1520 & ~ & 1518 & ~ & 1490 & ~ & ~ & 2513 & ~ & 1561 & ~ & 1511 & ~ & 1505 & ~ & ~ \\ \hline
        ~ & ~ & \multicolumn{9}{c|}{Utilization 95\%} & \multicolumn{9}{c|}{Utilization 98\%} \\ \hline
        Type & Variation & \multicolumn{2}{c}{MRP} & \multicolumn{2}{c}{Deterministic} & \multicolumn{2}{c}{Stochastic} & \multicolumn{2}{c|}{Stochastic} & \%$\Delta$ & \multicolumn{2}{c}{MRP} & \multicolumn{2}{c}{Deterministic} & \multicolumn{2}{c}{Stochastic} & \multicolumn{2}{c|}{Stochastic} & \%$\Delta$\\
        ~ & ~ & \multicolumn{2}{c}{(LT,SS,LSP)} & \multicolumn{2}{c}{(LT,SS)} & \multicolumn{2}{c}{($\tilde{T}$:12) (LT,SS)}& \multicolumn{2}{c|}{($\tilde{T}$:4) (LT,SS)}& &  \multicolumn{2}{c}{(LT,SS,LSP)} & \multicolumn{2}{c}{(LT,SS)} & \multicolumn{2}{c}{($\tilde{T}$:12) (LT,SS)} & \multicolumn{2}{c|}{($\tilde{T}$:1) (LT,SS)} &\\ \hline
        A\textsuperscript{1} & 2.5\% & 1138 & (1,0,P: 1) & \textbf{1134} & (1,0) & \multicolumn{2}{c}{-} & \multicolumn{2}{c|}{-} & 0\% & 1126 & (1,0,P: 1) & \textbf{1117} & (1,0) & \multicolumn{2}{c}{-} & \multicolumn{2}{c|}{-} & -1\% \\ 
        A\textsuperscript{1} & 5.0\% & 1142 & (1,0,P: 1) & \textbf{1135} & (1,0) & \multicolumn{2}{c}{-} & \multicolumn{2}{c|}{-} & -1\% & 1220 & (1,0,P: 1) & \textbf{1145} & (1,0) & \multicolumn{2}{c}{-} & \multicolumn{2}{c|}{-} & -6\% \\ 
        A\textsuperscript{1} & 7.5\% & 1175 & (1,0,P: 1) & \textbf{1143} & (1,0) & \multicolumn{2}{c}{-} & \multicolumn{2}{c|}{-} & -3\% & 1443 & (1,0,P: 1) & \textbf{1282} & (1,0) & \multicolumn{2}{c}{-} & \multicolumn{2}{c|}{-} & -11\% \\ 
        A\textsuperscript{1} & 10.0\% & 1246 & (1,0,P: 1) & \textbf{1157} & (1,0) & \multicolumn{2}{c}{-} & \multicolumn{2}{c|}{-} & -7\% & 1837 & (1,0,P: 1) & \textbf{1416} & (1,0) & \multicolumn{2}{c}{-} & \multicolumn{2}{c|}{-} & -23\% \\ 
        A\textsuperscript{1} & 12.5\% & 1362 & (1,0,P: 1) & \textbf{1187} & (1,0) & \multicolumn{2}{c}{-} & \multicolumn{2}{c|}{-} & -13\% & 2411 & (1,0,P: 1) & \textbf{1521} & (1,0) & \multicolumn{2}{c}{-} & \multicolumn{2}{c|}{-} & -37\% \\ \hline
        B\textsuperscript{2} & 2.5\% & 2591 & (2,0,P: 1) & \textbf{1264} & (1,0.1) & 1339 & (1,0.1) & 1339 & (1,0.1) & -51\% & 2578 & (2,0,P: 1) & 1319 & (1,0.1) & 1322 & (1,0.1) & \textbf{1293} & (1,0) & -49\% \\ 
        B\textsuperscript{2} & 5.0\% & 3313 & (2,0,Q: 1) & \textbf{1427} & (1,0.2) & 1432 & (1,0.1) & 1435 & (1,0.1) & -57\% & 3252 & (2,0,Q: 1) & 1657 & (1,0.3) & \textbf{1514} & (1,0.1) & 1515 & (1,0.1) & -53\% \\ 
        B\textsuperscript{2} & 7.5\% & 3274 & (2,0,Q: 1) & 1641 & (1,0.3) & \textbf{1561} & (1,0.1) & 1565 & (1,0.1) & -52\% & 3451 & (2,0,Q: 1.25) & 2061 & (1,0.4) & \textbf{1768} & (1,0.1) & 1776 & (1,0.1) & -49\% \\ 
        B\textsuperscript{2} & 10.0\% & 3464 & (1,0.6,Q: 1.25) & 1906 & (1,0.4) & \textbf{1723} & (1,0.1) & 1733 & (1,0.1) & -50\% & 3457 & (2,0,Q: 1.25) & 2352 & (1,0.5) & \textbf{1913} & (1,0.2) & 1941 & (1,0.1) & -45\% \\ 
        B\textsuperscript{2} & 12.5\% & 3537 & (2,0,Q: 1.25) & 2218 & (1,0.6) & 1917 & (1,0.2) & \textbf{1901} & (1,0.2) & -46\% & 3501 & (2,0,Q: 1.25) & 2609 & (1,0.6) & \textbf{2098} & (1,0.2) & 2145 & (1,0.2) & -40\% \\ \hline
        C\textsuperscript{3} & 2.5\% & 3306 & (1,0.6,Q: 1.25) & 1424 & (1,0.2) & \textbf{1347} & (1,0.1) & \textbf{1347} & (1,0.1) & -59\% & 3370 & (2,0.1,P: 2) & 1739 & (1,0.3) & \textbf{1559} & (1,0.1) & \textbf{1559} & (1,0.1) & -54\% \\ 
        C\textsuperscript{3} & 5.0\% & 3482 & (2,0,Q: 1.25) & 1974 & (1,0.4) & 1688 & (1,0.1) & \textbf{1677} & (1,0.2) & -52\% & 3415 & (2,0,Q: 1.25) & 2387 & (1,0.6) & \textbf{1873} & (1,0.2) & 1954 & (1,0.2) & -45\% \\ 
        C\textsuperscript{3} & 7.5\% & 3544 & (2,0,Q: 1.25) & 2419 & (1,0.6) & 1989 & (1,0.2) & \textbf{1952} & (1,0.2) & -44\% & 3620 & (2,0,Q: 1.5) & 2807 & (1,0.6) & \textbf{2195} & (1,0.2) & 2358 & (1,0.2) & -39\% \\ 
        C\textsuperscript{3} & 10.0\% & 3736 & (2,0,Q: 1.5) & 2842 & (1,0.6) & 2303 & (1,0.2) & \textbf{2242} & (1,0.2) & -38\% & 3969 & (2,0,Q: 1.75) & 3320 & (1,0.6) & \textbf{2643} & (1,0.2) & 2862 & (1,0.5) & -34\% \\ 
        C\textsuperscript{3} & 12.5\% & 3988 & (2,0,Q: 1.75) & 3317 & (1,0.6) & 2567 & (1,0.3) & \textbf{2564} & (1,0.4) & -35\% & 4352 & (2,0.1,Q: 2) & 4122 & (2,0.5) & \textbf{2907} & (1,0.4) & 3202 & (1,0.5) & -33\% \\ \hline
        Average & ~ & 2687 & ~ & 1746 & ~ & 1575 & ~ & 1593 & ~ & ~ & 2867 & ~ & 2057 & ~ & 1750 & ~ & 1816 & ~ & ~ \\ \hline
        \multicolumn{20}{|l|}{\textsuperscript{1}no changes, \textsuperscript{2}change at due date, \textsuperscript{3}change every period} \\
        \multicolumn{20}{|l|}{LSP = lot size policy, LT = lead time, P = fixed order period, Q = fixed order quantity, SS = safety stock, \%$\Delta$ = cost reduction to MRP} \\ \hline
    \end{tabular}
    }
    \caption{General cost and parameter overview across different resource utilizations, customer demand update patterns and demand variations. Comparison of MRP (LT,SS,LSP), deterministic optimization (LT,SS) and stochastic optimization (LT,SS) (for 30 scenarios)}
    \label{tab:general_overview_combined_table}
\end{sidewaystable}

\clearpage
\section{Cumulative flow diagrams}
\label{sec:Cumulativ_flow_Diagrams_Appendix}

\begin{figure}[ht]
    \centering
    \includegraphics[width=0.8\textwidth]{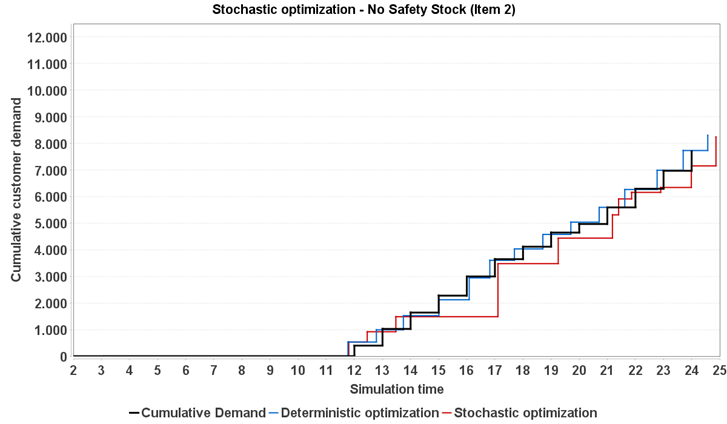}
    \caption{Cumulative Flow Diagram: Comparison of deterministic vs. stochastic optimization (Item 2, No Safety Stock, 90\% utilization, $\tilde{T}=12$)}
    \label{fig:cumulative_flow_diagramm_no_safety_stock}
\end{figure}

\begin{figure}[ht]
    \centering
    \includegraphics[width=0.8\textwidth]{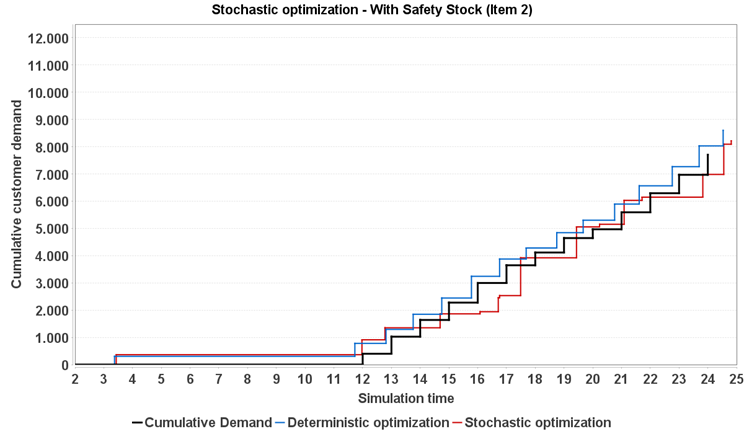}
    \caption{Cumulative Flow Diagram: Comparison of deterministic vs. stochastic optimization (Item 2, With Safety Stock, 90\% utilization, $\tilde{T}=12$)}
    \label{fig:cumulative_flow_diagramm_with_safety_stock}
\end{figure}

\end{document}